\documentclass[preprint,sort,compress,12pt]{elsarticle}

\usepackage{amssymb}
\usepackage{amsthm}
\usepackage{amsmath}
\usepackage{mathtools}
\usepackage{mathrsfs}
\usepackage{algorithm}
\usepackage[algo2e]{algorithm2e}
\usepackage{algpseudocode}
\usepackage{array}
\usepackage{multirow}
\usepackage{listings}
\usepackage{tabu}
\usepackage{booktabs}
\usepackage{enumerate}
\usepackage{fullpage}
\usepackage{float}
\usepackage{xcolor}
\usepackage[colorinlistoftodos]{todonotes}
\usepackage{bbm}
\usepackage{bm}
\usepackage[colorlinks=true]{hyperref}
\usepackage{url}
\usepackage{textcomp}
\usepackage{gensymb}
\usepackage{soul}
\usepackage{lineno}
\usepackage{graphicx}
\usepackage{subfigure}
\usepackage{float}
\usepackage{caption}
\usepackage{tikz}
\usetikzlibrary{shapes}

\theoremstyle{definition}

\theoremstyle{remark}

\biboptions{numbers,comma,round,square}
\graphicspath{ {./figs/} }

\journal{Elsevier}

\begin{document}
\begin{frontmatter}

\title{Generative Reconstruction of Spatiotemporal Wall-Pressure in Turbulent Boundary Layers via Patchwise Latent Diffusion}



\author[cornellMAE]{Xiantao Fan}
\author[cornellMAE]{Meet Hemant Parikh}
\author[cornellMAE,ndAME]{Yi Liu}
\author[ndAME]{Xin-Yang Liu}
\author[cornellMAE]{Junyi Guo}
\author[ndAME]{Meng Wang}
\author[cornellMAE,ndAME]{Jian-Xun Wang\corref{corxh}}

\address[cornellMAE]{Sibley School of Mechanical and Aerospace Engineering, Cornell University, Ithaca, NY, USA}
\address[ndAME]{Department of Aerospace and Mechanical Engineering, University of Notre Dame, Notre Dame, IN}

\cortext[corxh]{Corresponding author. Tel: +1 540 315 6512}
\ead{jw2837@cornell.edu}

\begin{abstract}

Wall-pressure fluctuations in turbulent boundary layers drive flow-induced noise, structural vibration, and hydroacoustic disturbances, especially in underwater and aerospace systems. Accurate prediction of their wavenumber-frequency spectra is critical for mitigation and design, yet empirical/analytical models rely on simplifying assumptions and miss the full spatiotemporal complexity, while high-fidelity simulations are prohibitive at high Reynolds numbers. Experimental measurements, though accessible, typically provide only pointwise signals and lack the resolution to recover full spatiotemporal fields. We propose a probabilistic generative framework that couples a patchwise (domain-decomposed) conditional neural field with a latent diffusion model to synthesize spatiotemporal wall-pressure fields under varying pressure-gradient conditions. The model conditions on sparse surface-sensor measurements and a low-cost mean-pressure descriptor, supports zero-shot adaptation to new sensor layouts, and produces ensembles with calibrated uncertainty. Validation against reference data shows accurate recovery of instantaneous fields and key statistics.

\end{abstract}

\end{frontmatter}


\section{Introduction}
\label{sec:intro}

Wall-pressure fluctuations beneath turbulent boundary layers (TBLs) are central to many engineering problems. They drive flow-induced structural vibration and aeroacoustics, contributing to fatigue, noise pollution, and loss of integrity in systems from aircraft and submarines to trains and automobiles~\cite{na1998structure, devenport2018sound, lee2021turbulent}. Accurate characterization of spatiotemporal wall-pressure fluctuations and its wavenumber-frequency spectra is essential for providing inputs to structural and acoustic models that predict flow-induced vibration and noise~\cite{wang2006computational, yang2022wavenumber}. 

Research over the past several decades has investigated wall-pressure fluctuations through experiments, high-fidelity simulations, and empirical/analytical models~\cite{grasso2019analytical}. Experiments deliver high-fidelity point signals, but the measured spectra are convolved with the device transfer function, making deconvolution and multiscale recovery difficult~\cite{prigent2020deconvolution}; moreover, such measurements are often sparse and cannot resolve the large-scale, spatially coherent structures of TBLs. High-fidelity numerical approaches such as direct numerical simulation (DNS) or wall-resolved large-eddy simulation (WRLES) can resolve wall-pressure dynamics and near-wall turbulence, but their computational cost increases rapidly with Reynolds number, making them impractical for design or optimal control applications~\cite{choi2012grid}. As a faster alternative, analytical and empirical models directly estimate wall-pressure power spectral densities (PSDs), often via the Poisson equation in the wavenumber-frequency domain or by normalizing canonical spectra~\cite{slama2018kriging, pereira2022physics, panton1974wall}. However, these approaches require accurate high-order turbulence statistics (e.g., Reynolds stress) as input, which are difficult to obtain in practice. And they generalize poorly beyond zero-pressure-gradient (ZPG) flows, especially under adverse/favorable pressure gradients (APG/FPG)~\cite{thomson2021comparison, fritsch2023modeling}. Critically, PSD-based models yield spectra rather than instantaneous fields and therefore cannot provide the spatiotemporal fluctuations needed for flow control. As a result, robust reconstruction of wall-pressure fields and their spectra across non-equilibrium boundary layers remains challenging.

To tackle these challenges, machine learning (ML) offers a complementary route. ML can learn expressive nonlinear mappings, enabling flexible modeling of high-dimensional turbulence. One prominent line develops spectrum-level surrogates: for example, gene expression programming (GEP) has been used to infer analytical formulas for pointwise frequency spectra and adapt scaling to pressure-gradient effects, especially improving low-frequency predictions in APG flows~\cite{dominique2021inferring, fritsch2023modeling, shubham2023data, shubham2024more, jin2025assessment, bolavar2025empirical}. Lightweight neural networks likewise fit spectra in a black-box manner with higher accuracy and training efficiency than classical curve fitting~\cite{kurhade2023artificial, dominique2022artificial, zeng2025rapid}. However, these approaches are inherently narrow in scope: they are typically trained for specific conditions and degrade when extrapolated; more fundamentally, by operating on pointwise, one-dimensional spectra, they discard spatial phase and cross-spectral coherence. As a result, they provide limited insight into instantaneous physics and are inadequate for tasks that require full spatiotemporal wall-pressure reconstruction.

In response to these gaps, growing effort has focused on fusing wall-mounted sensor data with ML to reconstruct spatially and temporally resolved wall-pressure fields. Wall sensors are easy to deploy and provide time-resolved, non-intrusive measurements at scale. Leveraging such data, trained models can infer full fields and subsequently yield wavenumber-frequency spectra and other turbulence statistics via post-processing. For example, Poulinakis et al.~\cite{poulinakis2023deep} combined a long short-term memory (LSTM) network with cubic-spline interpolation to upsample coarse pressure signals, accurately recovering low-frequency content. More broadly, ML has been widely applied to flow reconstruction, especially super-resolution of \emph{velocity} fields using convolutional neural networks (CNNs) on structured grids~\cite{fukami2019super, fukami2023super, fukami2022machine} and graph neural networks on unstructured meshes~\cite{danciu2024flow}. Recent advances incorporate skip connections~\cite{sofos2024comparison}, attention mechanisms~\cite{cheng2025improved, zeng2024super}, proper orthogonal decomposition (POD)-informed models~\cite{diop2022reconstruction}, and mesh-aware CNNs with spatially varying filters~\cite{hu2022mesh}; some works incorporate physics-informed losses to promote physical consistency during training~\cite{yadav2025rf, zhu2024new}. Despite this progress, most ML reconstructions are fundamentally \emph{deterministic}. While such models perform well for relatively simple flows~\cite{fukami2019super}, including laminar regimes, two-dimensional vorticity dynamics, and jet flows~\cite{lee2024superresolution}, they struggle to capture the inherent stochasticity of fully turbulent wall-pressure fields. Moreover, reconstructing full fields from sparse sensors is a severely ill-posed inverse problem: multiple plausible fields fit the same data. Deterministic ML yields a single point estimate that cannot represent posterior uncertainty, often blurs coherent structures, and underestimates extreme events. These limitations motivate the development of \emph{probabilistic} generative frameworks capable of capturing the inherent uncertainty while preserving statistical stationarity in realistic wall-pressure dynamics.

Among generative modeling frameworks, diffusion models have emerged as state-of-the-art for synthesizing high-fidelity flow realizations that reproduce target turbulence statistics. Recent studies demonstrate diffusion for flow generation, reconstruction, and super-resolution~\cite{ruhling2023dyffusion,molinaro2024generative,shu2023physics,kohl2023benchmarking,gao2024bayesian,shehataimproved,du2024conditional,liu2024confild,fan2025neural}. A notable exemplar is the conditional neural field-based latent diffusion framework (CoNFiLD), which synthesizes spatiotemporal (4D) wall-bounded turbulence while preserving inhomogeneity and anisotropy, and enables synthetic inflows and zero-shot sparse-sensor reconstruction~\cite{du2024conditional,liu2024confild}. In a Bayesian posterior-sampling view, such trained generative priors can act as \emph{foundation models} that support training-free reconstruction across diverse sensor arrangements, offering a promising route to flow reconstruction conditioned on sparse wall-pressure sensors regardless of placement or count~\cite{hemant2025conditional}. However, convection-dominated \emph{wall-pressure} signals pose additional challenges: the elliptic coupling of the pressure Poisson equation induces long-range, nonlocal dependencies and complex space-time coherence. To our knowledge, no prior work reconstructs full spatiotemporal wall-pressure fields from sparse wall sensors using a generative model. 

In this work, we propose a task-specific generative framework for wall-pressure reconstruction in TBLs that combines (i) a \emph{domain-decomposed conditional neural field} producing compact local latents with high-resolution decoding and (ii) a \emph{patched latent diffusion} model equipped with classifier-free \emph{regime guidance} and diffusion posterior sampling for training-free data assimilation. The resulting model adapts at inference to arbitrary sparse sensor layouts without retraining, reconstructs spatiotemporal wall-pressure fields with calibrated uncertainty, and generalizes across a range of APG/FPG conditions. 
The remainder of the paper is organized as follows. Section~\ref{sec:meth} details the proposed framework. Section~\ref{sec:res} presents reconstruction results across different flow conditions to demonstrate the effectiveness of the approach. Section~\ref{sec:dis} analyzes generalization and limitations. Section~\ref{sec:conclusion} concludes and outlines future directions.

\section{Methodology}
\label{sec:meth}

\subsection{Problem formulation} 
\label{sec:problem equation}

We consider an incompressible turbulent boundary layer (TBL) over a smooth, rigid wall. Using Reynolds decomposition $u_i = U_i + u_i'$ and $p = \overline{p}+p'$, the fluctuating pressure field $p'(\mathbf{x},t)$ satisfies the Poisson equation obtained by taking the divergence of the momentum equations~\cite{grasso2019analytical}:
\begin{equation}
-\frac{\partial^2 p'}{\partial x_i^2} = 2\rho\frac{\partial U_i}{\partial x_j}\frac{\partial u'_j}{\partial x_i} + \rho \frac{\partial^2}{\partial x_i \partial x_j}(u'_iu'_j - \overline{u'_iu'_j})
\label{eq:poisson}
\end{equation}
where $\rho$ is the density; $x_i$ denotes the Cartesian coordinates; $U_i$ and $u'_i$ are the mean and fluctuating velocity components, respectively; $i,j\in\{1,2,3\}$ correspond to the streamwise, wall-normal, and spanwise directions; and repeated indices imply summation. The two right-hand-side terms are the classical \emph{rapid} (mean-shear-turbulence) and \emph{slow} (turbulence-turbulence) contributions.  Although our target is the wall-pressure fluctuation field, the elliptic character of Eq.~\eqref{eq:poisson} implies strong nonlocality: local pressure depends on the global distribution of velocity gradients and Reynolds stresses.

We define the wall-pressure fluctuation $p_w(x,z,t) \equiv p'(x,0,z,t)$ on the wall plane. In principle, one could obtain $p_w$ by solving Eq.~\eqref{eq:poisson} (e.g., via Green's functions). In practice, this would require accurate Reynolds stresses and velocity fluctuations throughout the near-wall volume, which are notoriously difficult to measure or numerically resolve at high Reynolds numbers. By contrast, wall pressure can be measured reliably and nonintrusively with flush-mounted sensors. However, such measurements are often \emph{sparse} (limited mounting locations) and \emph{noisy} (finite sampling rate and sensor noise), so inferring the full, densely resolved and globally coupled field from these measurements is underdetermined.

\begin{figure}[t!]
    \centering
    \includegraphics[width=\textwidth]{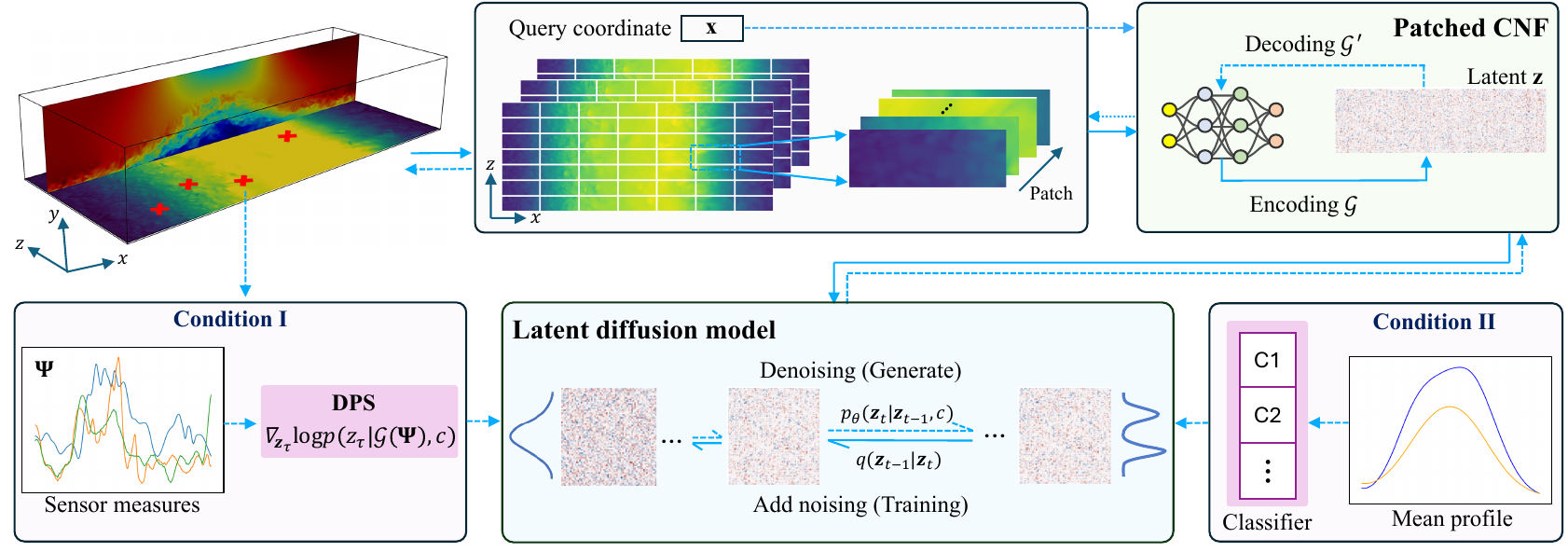}
    \caption{Schematic of the generative framework for reconstructing wall pressure. \emph{Top:} the domain-decomposed conditional neural field (D-CNF) encodes the physical domain into latent space and decodes latents back to fields. \emph{Bottom:} the latent diffusion model with two conditioning mechanisms: (I) sensor consistency via DPS (inference only) and (II) flow-regime guidance via a mean-profile descriptor used in both training and inference. Solid arrows denote training; dashed arrows denote inference.}
    \label{fig:schematic}
\end{figure}

As introduced in Sec.~\ref{sec:intro}, most ML approaches for wall pressure target direct prediction of spectral quantities~\cite{meloni2023experimental,shubham2023data,kurhade2023artificial,fritsch2023modeling,dominique2021inferring}. In contrast, we aim to reconstruct the full spatiotemporal wall-pressure field from sparse sensors with \emph{calibrated uncertainty}. Because the inverse map is many-to-one and the dynamics are stochastic, a generative formulation is appropriate. Following the representation-dynamics factorization of CoNFiLD~\cite{du2024conditional}, our framework (Fig.~\ref{fig:schematic}) first uses a \emph{domain-decomposed conditional neural field} (D-CNF)~\cite{guo2025conditional} to encode each instantaneous wall-pressure field into a compact latent $z_t$, with a decoder that maps latents back to the physical domain. In the latent space, a \emph{classifier-free patched diffusion model} learns the stochastic temporal evolution. Patching induces locality on the wall plane and scales to long spans and high resolutions. Two conditioning signals steer generation toward full-field reconstructions. The first is \emph{sensor consistency} enforced via diffusion posterior sampling (DPS) during inference only: given sparse measurements, the reverse diffusion is augmented by the sensor likelihood gradient backpropagated through the D-CNF decoder, enabling agreement with arbitrary sensor layouts without retraining. The second is \emph{flow-regime guidance} provided by a low-cost descriptor $c$ (e.g., a mean wall-pressure profile obtainable from a RANS simulation). Using classifier-free guidance, $c$ is applied during both training and inference, which promotes zero-shot generalization across different adverse/favorable pressure-gradient (APG/FPG) flow regimes. With DPS and regime guidance, the diffusion model generates latent trajectories that are decoded to produce measurement-aligned spatiotemporal reconstructions of instantaneous $p_w(x,z,t)$, while preserving physically relevant statistics (e.g., wavenumber-frequency spectra and coherence) for downstream applications.

\subsection{Domain-decomposed conditional neural field for latent representation}
\label{sec:CNF}

The first component is a \emph{domain-decomposed conditional neural field} (D-CNF) that maps each instantaneous wall-pressure field to a compact latent code and decodes latents back to the physical domain. We adopt an \emph{auto-decoder} design with a \emph{fully projected} hypernetwork, following the CoNFiLD line~\cite{liu2024confild} and recent non-overlapping CNF tiling study~\cite{guo2025conditional}. In an auto-decoder, no amortized encoder is trained; instead, each training sample has a learnable latent code, while at inference latents are supplied by the generative model.

Let the wall plane be $\Omega_w\subset\mathbb{R}^2$ with in-plane coordinates $\mathbf{x}=(x,z)$. We partition $\Omega_w$ into $N_p$ non-overlapping tiles $\{\Omega_i\}_{i=1}^{N_p}$ whose interiors are disjoint and union (up to boundaries) equals $\Omega_w$. For each tile, an affine map $\mathcal{T}_i:\Omega_i\to[-1,1]^2$ normalizes coordinates to promote parameter sharing. We denote by $t$ the discrete time index and by $c\in\mathbb{R}^{d_c}$ a flow-regime descriptor (e.g., Clauser parameter $\beta$, or a mean wall-pressure/profile summary from a low-cost RANS computation). The D-CNF predicts per-tile fields $\hat p_{w,i}(\mathbf{x},t,c)$ and assembles the global reconstruction by tiling: $\hat p_w(\mathbf{x},t,c) = \hat p_{w,i}(\mathbf{x},t,c), \mathbf{x}\in\Omega_i,i=1,\dots,N_p$, so no explicit blending is required and locality is preserved. Each tile is modeled by a sinusoidal representation network (SIREN)-based implicit field $\mathcal{F}_\theta$~\cite{sitzmann2020implicit} acting on local coordinates $\tilde{\mathbf{x}}=\mathcal{T}_i(\mathbf{x})\in[-1,1]^2$:
\begin{equation}
\hat p_{w,i}(\mathbf{x},t,k) = \mathcal{F}_{\theta_{i,t,c}} \big(\tilde{\mathbf{x}}\big),
\qquad \theta_{i,t,c} = \zeta + \mathcal{H}_\gamma \big(\mathbf{s}_{i,t,c}, t, c\big),
\label{eq:cnf_forward}
\end{equation}
where $\mathbf{s}_{i,t,c}\in\mathbb{R}^{d_z}$ is the learnable latent for tile $i$ at time $t$ and descriptor $c$; $\zeta$ collects the shared base SIREN parameters; and $\mathcal{H}_\gamma$ is a \emph{fully projected} hypernetwork (parameters $\gamma$) that outputs layer-wise weight/bias offsets applied to all SIREN layers. This projection is more expressive than FiLM-style modulation and has proved effective for convection-dominated wall-pressure structure~\cite{guo2025conditional}.

Let $\mathcal{D}=\{(t_j,c_j,p_w(\cdot,t_j,c_j))\}_{j=1}^{N_s}$ be the training set of snapshots with continuous descriptors. For each sample $j$ and tile $i$, we associate a latent $\mathbf{s}_{i,j}$ (equivalently $\mathbf{s}_{i,t_j,c_j}$). Denote the restriction of $p_w$ to $\Omega_i$ by $p_{w,i}(\mathbf{x};t_j,c_j)$. We jointly optimize latents and network parameters by minimizing an $L_2$ reconstruction loss over points sampled in each tile:
\begin{equation}
(\mathbf{s}^*,\zeta^*,\gamma^*)
=
\arg\min_{\mathbf{L},\zeta,\gamma}
\sum_{j=1}^{N_s}\sum_{i=1}^{N_p}
\left\|
p_{w,i}(\mathbf{x};t_j,c_j)
-
\mathcal{F}_{\zeta+\mathcal{H}_\gamma(\mathbf{z}_{i,j},\, t_j,\, c_j)}\!\big(\mathcal{T}_i(\mathbf{x})\big)
\right\|_{L_2(\Omega_i)}^2.
\label{eq:cnf_loss}
\end{equation}
A small $\ell_2$ regularizer on $\mathbf{s}_{i,j}$ may be used for stability, though we did not find it necessary.
The SIREN backbone uses the standard input-frequency scale $\omega_0$ controlled by $s_0$ and initialization $W \sim \mathcal{U}(-r/\sqrt{n},\,r/\sqrt{n})$, where $\mathcal{U}(\cdot)$ represents a uniform distribution with specified bounds~\cite{sitzmann2020implicit}. Typical choices $s_0=30$ and $r=\sqrt{6}$ were robust in our experiments. Full architectural details are provided in~\ref{sec: arch_para_pcnf}.
The D-CNF therefore yields per-tile latents that (i) capture localized, convection-dominated features, (ii) provide exact gradients $\partial \hat p/\partial \mathbf{s}$ for sensor-consistency in DPS, and (iii) decode to high-resolution fields via the non-overlapping tiling.

\subsection{Conditional latent diffusion model with classifier-free guidance}
\label{sec:method_diff}

At physical time index $t$, the D-CNF (Sec.~\ref{sec:CNF}) yields per-tile latents $\{\mathbf{s}_{i,t,c} \in \mathbb{R}^{d_z}\}_{i=1}^{N_p}$ for a given flow regime descriptor $c\in\mathbb{R}^{d_c}$. For temporal modeling we form a short window of length $T_w$ and, for each tile $i$, stack the latents into a spatiotemporal tensor $\mathbf{z}_0^{(i)} \in\mathbb{R}^{T_w\times d_z}$. Diffusion is performed on $\mathbf{z}_\tau^{(i)}$ with tiles processed in parallel as the batch dimension. For notational brevity below we drop the tile index and write $\mathbf{z}_\tau\in\mathbb{R}^{T_w\times d_z}$.

With a variance schedule $\{\beta_\tau\}_{\tau=1}^S$, $\alpha_\tau=1-\beta_\tau$, and $\bar\alpha_\tau=\prod_{u=1}^\tau\alpha_u$, the forward (noising) process is defined by,
\begin{equation}
q \left(\mathbf{z}_\tau \mid \mathbf{z}_{\tau-1}\right)
=\mathcal{N}\big(\sqrt{\alpha_\tau}\mathbf{z}_{\tau-1},\beta_\tau \mathbf{I}\big),
\qquad
q \left(\mathbf{z}_\tau \mid \mathbf{z}_0\right)
=\mathcal{N}\big(\sqrt{\bar\alpha_\tau}\,\mathbf{z}_0,(1-\bar\alpha_\tau)\mathbf{I}\big),
\label{eq:forward_ddpm}
\end{equation}
where $\tau\in\{1,\dots,S\}$ indexes diffusion steps (distinct from physical time $t$). 

The reverse process (generation) is achieved by learning a noise predictor $\boldsymbol{\epsilon}_\theta(\mathbf{z}_\tau,\tau,\tilde c)$ with parameters $\theta$, where $\tilde c$ is either the descriptor $c$ or a null token $\emptyset$ (for classifier-free learning). The reverse transition is parameterized as
\begin{equation}
p_\theta\left(\mathbf{z}_{\tau-1}\mid \mathbf{z}_\tau, c\right)
=\mathcal{N}\big(\boldsymbol{\mu}_\theta(\mathbf{z}_\tau,\tau,\tilde{c}),\tilde\beta_\tau \mathbf{I}\big),\quad
\boldsymbol{\mu}_\theta(\mathbf{z}_\tau,\tau,\tilde{c})
=\frac{1}{\sqrt{\alpha_\tau}}\!\left(\mathbf{z}_\tau-\frac{\beta_\tau}{\sqrt{1-\bar\alpha_\tau}}
\boldsymbol{\epsilon}_\theta(\mathbf{z}_\tau,\tau,\tilde{c})\right),
\label{eq:reverse_gaussian_std}
\end{equation}
with $\tilde\beta_\tau$ following the usual denoising diffusion probabilistic model (DDPM) choice. The training is performed using the simplified noise-matching loss,
\begin{equation}
\mathcal{L}_{\text{diff}}
=\mathbb{E}_{\mathbf{z}_0,\,\tau\sim\mathcal{U}[1,S],\,\boldsymbol{\epsilon}\sim\mathcal{N}(0,\mathbf{I}),\,\tilde c}
\left[
\big\|\boldsymbol{\epsilon}-\boldsymbol{\epsilon}_\theta\big(\sqrt{\bar\alpha_\tau}\mathbf{z}_0+\sqrt{1-\bar\alpha_\tau}\,\boldsymbol{\epsilon},\, \tau,\, \tilde c\big)\big\|_2^2
\right].
\label{eq:diff_loss}
\end{equation}

To generalize across a range of flow conditions with varying pressure gradients, we adopt a classifier-free guidance (CFG) strategy to handle the flow descriptor $c$ without training a separate noisy-latent classifier and to avoid the diversity collapse often induced by classifier guidance~\cite{ho2022classifier}. CFG trains a single noise predictor on both conditional and unconditional inputs by randomly dropping $c$ with probability $p_{\mathrm{uncond}}$, then steers sampling with a tunable guidance strength $w$ while preserving stochastic variability. During training we set $\tilde c=\emptyset$ with probability $p_{\mathrm{uncond}}$, and at inference we combine scores as
\begin{equation}
\tilde{\boldsymbol{\epsilon}}_\theta(\mathbf{z}_\tau,\tau,c)
=(1+w)\,\boldsymbol{\epsilon}_\theta(\mathbf{z}_\tau,\tau,c)
- w\,\boldsymbol{\epsilon}_\theta(\mathbf{z}_\tau,\tau,\emptyset),
\label{eq:cfg_score}
\end{equation}
where $w$ controls guidance ($w=-1$ unconditional, $w=0$ conditional only, $w>0$ extrapolated; we use $w=1$ for diverse pressure-gradient scenarios unless noted). The flow regime descriptor $c$ is embedded by a small multilayer perceptron (MLP) and fused with the timestep embedding at all U-Net blocks; we use $p_{\mathrm{uncond}}{=}0.2$.
Specifically, a modernized U-Net is used for $\boldsymbol{\epsilon}_\theta(\cdot)$ that operates on tensors of shape $T_w\times d_z$ (tiles in batch). The network employs residual convolutional blocks, group normalization, and multi-head self-attention over time to capture long-range spatiotemporal couplings in latent space. Diffusion step $\tau$ is encoded via sinusoidal embeddings; the descriptor embedding for $c$ is added to these embeddings at each block. More details of the network architecture are provided in ~\ref{sec: arch_para_diffusion}.

\subsection{Generative reconstruction from sparse wall measurements without retraining}

We aim to reconstruct the spatiotemporal wall-pressure field from sparse wall sensors under arbitrary layouts, \emph{without retraining}. As sensor number, placement, and sampling rate vary across cases, to let a single pre-trained model adapt zero-shot to any configuration, we assimilate sensor information \emph{exclusively at inference} via DPS~\cite{fan2025neural,gao2024bayesian,du2024conditional}. In DPS, the D-CNF and diffusion parameters are frozen; sensors influence sampling only through a likelihood gradient computed on the decoded field through the state-to-observable operator rather than by updating weights. 

Concretely, given the flow descriptor $c$ and sparse wall records $\mathbf{y}$ over a temporal window of length $T_w$, we sample the latent ``image'' $\mathbf{z}_0\in\mathbb{R}^{T_w\times d_z}$ using the DPS-modified reverse process. Classifier-free guidance provides the conditional prior score $\bm{s_\theta}(\mathbf{z}_\tau, \tau, c; \bm{\theta}^*)$,
\begin{equation}
    \bm{s_\theta}(\mathbf{z}_\tau, \tau, c; \bm{\theta}^*) = -\frac{\bm{\epsilon_\theta}(\mathbf{z}_\tau,  \tau, {c}; \bm{\theta}^*)}{\sqrt{(1-\bar{\alpha}_\tau)}}. 
\label{eq:score}
\end{equation}
The sensor measurements $\mathbf{y}$ are assimilated by modifying this to form a posterior score,
\begin{equation}
    \bm{s}_{\text{post}}(\mathbf{z}_\tau,\tau,c; \mathbf{y}, \bm{\theta}^*) 
    = \bm{s_\theta}(\mathbf{z}_\tau, \tau, c; \bm{\theta}^*) + \nabla_{\mathbf{z}_\tau}\log p(\mathbf{y}|\mathbf{z}_\tau),
\label{eq:dps_score}
\end{equation}
where the intractable marginal $p(\mathbf{y}|\mathbf{z}_\tau)$ is approximated by evaluating a pseudo-likelihood at a clean-latent estimate $\hat{\mathbf{z}}_0(\mathbf{z}_\tau)$~\cite{chung2023diffusion}. Using Tweedie's formula in the $\epsilon$-prediction parameterization~\cite{efron2011tweedie,kim2021noise2score},
\begin{equation}
\hat{\mathbf{z}}_0(\mathbf{z}_\tau)
\approx
\frac{1}{\sqrt{\bar{\alpha}_\tau}}
\Big(
\mathbf{z}_\tau + (1-\bar{\alpha}_\tau)\bm{s}_\theta(\mathbf{z}_\tau,\tau,c; \bm{\theta}^*)
\Big).
\label{eq:tweedie_clean}
\end{equation}
Assuming Gaussian measurement noise with scale $\sigma_c(\tau)$, the likelihood score can be approximated as,
\begin{equation}
\nabla_{\mathbf{z}_\tau}\log p(\mathbf{y}|\mathbf{z}_\tau) 
\simeq
\nabla_{\mathbf{z}_\tau}\log p(\mathbf{y}| 
\hat{\mathbf{z}}_0)
=
-\frac{1}{\sigma_c(\tau)^2}
\nabla_{\mathbf{z}_\tau}\frac{1}{2}
\big\|
\mathcal{M}(\mathcal{G}'(\hat{\mathbf{z}}_0))-\mathbf{y}
\big\|_2^2,
\label{eq:likelihood_grad}
\end{equation}
with the gradient backpropagated through $\mathbf{z}_\tau \rightarrow \hat{\mathbf{z}}_0(\mathbf{z}_\tau)\xrightarrow{\mathcal{G}'}$ field $\xrightarrow{\mathcal{M}}$ sensors. Following~\cite{chung2023diffusion}, we use $\sigma_c(\tau)=1$ unless stated; an annealed schedule can emphasize sensors late in the reverse process, where $\mathcal{G'}$ denotes the D-CNF decoder that maps the latent variable to the full pressure field, and $\mathcal{M}$ is spatial operator that extracts the sensor locations from the reconstructed field. Because $\mathbf{y}$ and $\mathcal{M}$ enter only through the likelihood, the same pre-trained model adapts \emph{zero-shot} to different sensor counts, layouts, and sampling rates, and no retraining or architectural changes are required. Each posterior sample $\mathbf{z}_0$ is finally decoded by $\mathcal{G}'$ to produce a coherent reconstruction; autoregressive temporal DPS yields long sequences~\cite{du2024conditional}, and ensembles provide posterior means and uncertainty.

\section{Results}
\label{sec:res}
\subsection{Case setup and training data generation}
\label{sec:dataset}

We construct a database of TBLs under varying pressure gradients using direct numerical simulations (DNS). The three-dimensional, incompressible Navier-Stokes equations are solved with a well-established finite-volume code developed at the Stanford University~\cite{you2008discrete}. The numerical setup follows Abe et al.~\cite{abe2017reynolds, abe2019direct}, where an artificial boundary condition is imposed at the upper boundary (blowing-suction) to mimic pressure-gradient effects, while a no-slip condition is applied at the wall. A two-dimensional separation bubble is induced in the $x$–$y$ plane; periodic boundary conditions are enforced in the spanwise ($z$) direction. In streamwise direction, inflow and convective outflow boundary conditions are employed. Turbulent inflow is supplied via the rescale-and-recycle procedure of Lund et al.~\cite{lund1998generation}. Unless noted, all quantities are normalized by outer variables (free-stream velocity $U_\infty$ and inlet momentum thickness $\theta_{\mathrm{in}}$). The Reynolds number based on the inlet momentum thickness is $Re_{\theta_{\mathrm{in}}}=300$.

\begin{figure}[t!]
    \centering
    \includegraphics[width=0.9\textwidth]{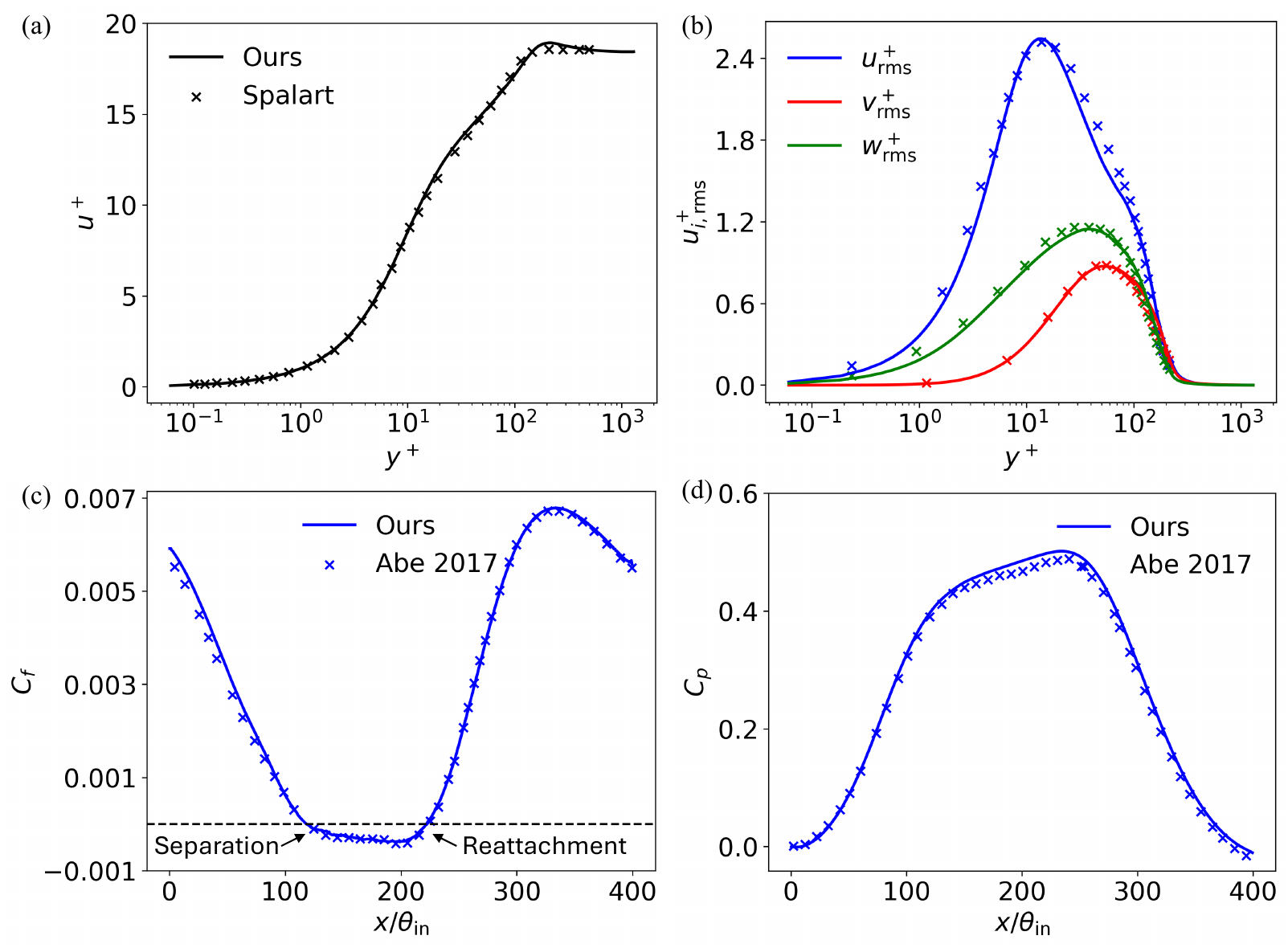}
    \caption{Dataset validation: (a) Mean velocity and (b) velocity fluctuations at the inlet, compared with DNS data from~\cite{spalart1988direct}; (c) Skin friction and (d) mean pressure coefficient for a turbulent boundary layer under strong APG/FPG (V90, $\beta = 90.19$), compared with DNS results from~\cite{abe2017reynolds}. Reference data are shown as scatter points, while our results are plotted as lines.}
    \label{fig:data_validation}
\end{figure}

\begin{table}[t!]
    \centering    
    \caption{Numerical settings and parameters for the DNS dataset}
    \footnotesize
    \begin{tabular}{c c c c c c c c}
    \toprule[1.5pt]
        Case & V0 & V20 & V40 & V60 & V80 & V90 & V100 \\
        \midrule
        $V_{\mathrm{max}}$ & 0 & 0.0665 & 0.133 & 0.1995 & 0.266 & 0.29925 & 0.3325\\
        $\beta_{x/\theta_{\mathrm{in}} = 100}$ & 0 & 2.14 & 6.92 & 22.31 & 62.31 & 90.19 &  302.55 \\
        Grid & \multicolumn{7}{c}{$N_x \times N_y \times N_z = 512 \times 320 \times 512$} \\
        Domain & \multicolumn{7}{c}{$L_x \times L_y \times L_z = 400 \times 80 \times 160$} \\
        $\Delta T$ & \multicolumn{7}{c}{$1.6$}\\
        $Re_{\theta_{\mathrm{in}}}$ & \multicolumn{7}{c}{$300$}\\
        $T^{\mathrm{train}}_{\mathrm{flow}} / N$ & \multicolumn{7}{c}{$0-4 / 1000$} \\
        $T^{\mathrm{test}}_{\mathrm{flow}} / N$ & \multicolumn{7}{c}{$8-12 / 1000$} \\
    \bottomrule[1.5pt] 
    \end{tabular}
    \label{tab:num_setting}
\end{table}
To enrich coverage beyond the three cases in~\cite{abe2017reynolds}, we vary the magnitude of the upper-boundary wall-normal blowing/suction to generate a wide range of TBL, including attached and separated flows, spanning zero-, adverse-, and favorable-pressure-gradient (ZPG, APG, FPG) regimes. The wall-normal velocity at the top boundary is prescribed as,
\begin{equation}
    V_{\mathrm{top}} = V_{\mathrm{max}} \sqrt{2} \Big(\frac{x_c - x}{\sigma}\Big) \exp\Big[\psi - \Big(\frac{x_c - x}{\sigma}\Big)^2 \Big] 
\label{eq:top_boundary}
\end{equation}
with $x_c = 195.3125$, $\sigma = 89$, and $\psi = 0.95$, and where $V_{\mathrm{max}}$ controls the pressure-gradient strength (see Tab.~\ref{tab:num_setting}). At the top boundary, the streamwise velocity $u$ is specified to maintain zero spanwise vorticity; a Neumann condition is imposed on the spanwise velocity $w$.  
We characterize pressure-gradient intensity via the Clauser parameter~\cite{abe2019direct},
\begin{equation}
\beta =\ \frac{\delta^{*}}{\tau_{w}}\,\frac{d\bar{p}}{dx},
\end{equation}
where $\delta^*$ is the displacement thickness and $\tau_w$ is the wall shear stress. For all cases, the largest pressure gradient occurs near $x/\theta_{\mathrm{in}} = 100$; therefore, we report the corresponding $\beta$ values at this location in Tab.~\ref{tab:num_setting}. 
Case labels V$\cdot$ denote the fraction of the reference amplitude (e.g., V20 $\equiv 0.2\,V_{\mathrm{max}}$; V100 $\equiv V_{\mathrm{max}}$). The suite spans ZPG, FPG, attached APG up to incipient separation, and strongly separated APG (V80-V100; $\beta>39$)~\cite{kitsios2017direct}. Numerical settings and $\beta$ values are also summarized in Tab.~\ref{tab:num_setting}.
We validate the DNS by comparing (i) inlet ZPG mean and fluctuation profiles against the reference DNS of Spalart~\cite{spalart1988direct}, and (ii) skin friction $C_f$ and mean pressure coefficient $C_p$ under representative APG/FPG (V90) against Abe~\cite{abe2017reynolds}. As shown in Fig.~\ref{fig:data_validation}, the present results agree closely with the references: the ZPG inner-outer profiles collapse onto the benchmark data, and the APG/FPG case reproduces the separation/reattachment behavior in $C_f$ and the streamwise variation of $C_p$. This agreement supports the fidelity of our simulations and their use for training and evaluation. The instantaneous streamwise velocity fields for all cases are presented in Fig.~\ref{fig:case_velocity} to illustrate the flow dynamics.

Wall-pressure snapshots on the wall plane have resolution $N_x\times N_z = 512\times 512$. Data are sampled every $\Delta T=1.6\theta_{\mathrm{in}}/U_\infty$ (roughly ten numerical time steps). Training covers four flow-through times ($T_{\mathrm{flow}}$); testing uses disjoint windows eight to twelve flow-through times downstream (Tab.~\ref{tab:num_setting}). Only cases V0, V40, and V80 are included in training; all other cases are held out for testing to assess cross-regime generalization. To ensure statistical independence between training and testing data, all test samples are extracted from time intervals beyond the initial four flow-through periods. Following Sec.~\ref{sec:CNF}, the wall plane is partitioned into $16\times 16$ non-overlapping tiles (256 total), each of size $32\times 32$, and encoded by the D-CNF into a latent vector of length $d_z=64$ per tile/time. For diffusion modeling, we form temporal windows of length $T_w$ and stack per-tile latents into spatiotemporal ``latent images'' of size $T_w\times d_z$ (tiles in the batch dimension). The resulting latent dataset $\{\mathbf{z}_0\}$ is used to train the diffusion model.

To test the reconstruction performance of the trained model, we consider two scenarios. \emph{Scenario I (single-regime testing):} the flow condition (e.g., $Re$ and pressure-gradient setting) is fixed, but the test sequences are \emph{new spatiotemporal realizations} drawn from disjoint time windows and/or new initial conditions; inference uses only sparse sensors (Condition I) via DPS, with no parameter conditioning. \emph{Scenario II (cross-regime generalization):} the model reconstructs fields at previously unseen pressure-gradient strengths (ZPG/APG/FPG). Training includes Condition II via classifier-free guidance with a continuous descriptor (e.g., mean $C_p$), and testing uses both sparse sensors (Condition I) and the descriptor (Condition II) to enable zero-shot transfer across regimes.

\subsection{Generative wall-pressure reconstruction (single-regime testing)}
\label{sec:cond_gen_sensor}

We first evaluate reconstruction in a \emph{single-regime testing} setting: the flow condition is fixed, but test sequences are new spatiotemporal realizations drawn from new initial conditions. Three representative regimes are considered, including ZPG (V0), moderate APG/FPG (V40), and strong APG/FPG with separation (V80). For each flow regime, a model is trained once and tested on new realizations. During inference, only Condition~I is used: wall-pressure measurements from a $16{\times}16$ sensor array co-located with the D-CNF tiles are assimilated via DPS.

\begin{figure}[H]
    \centering
    \includegraphics[width=0.9\textwidth]{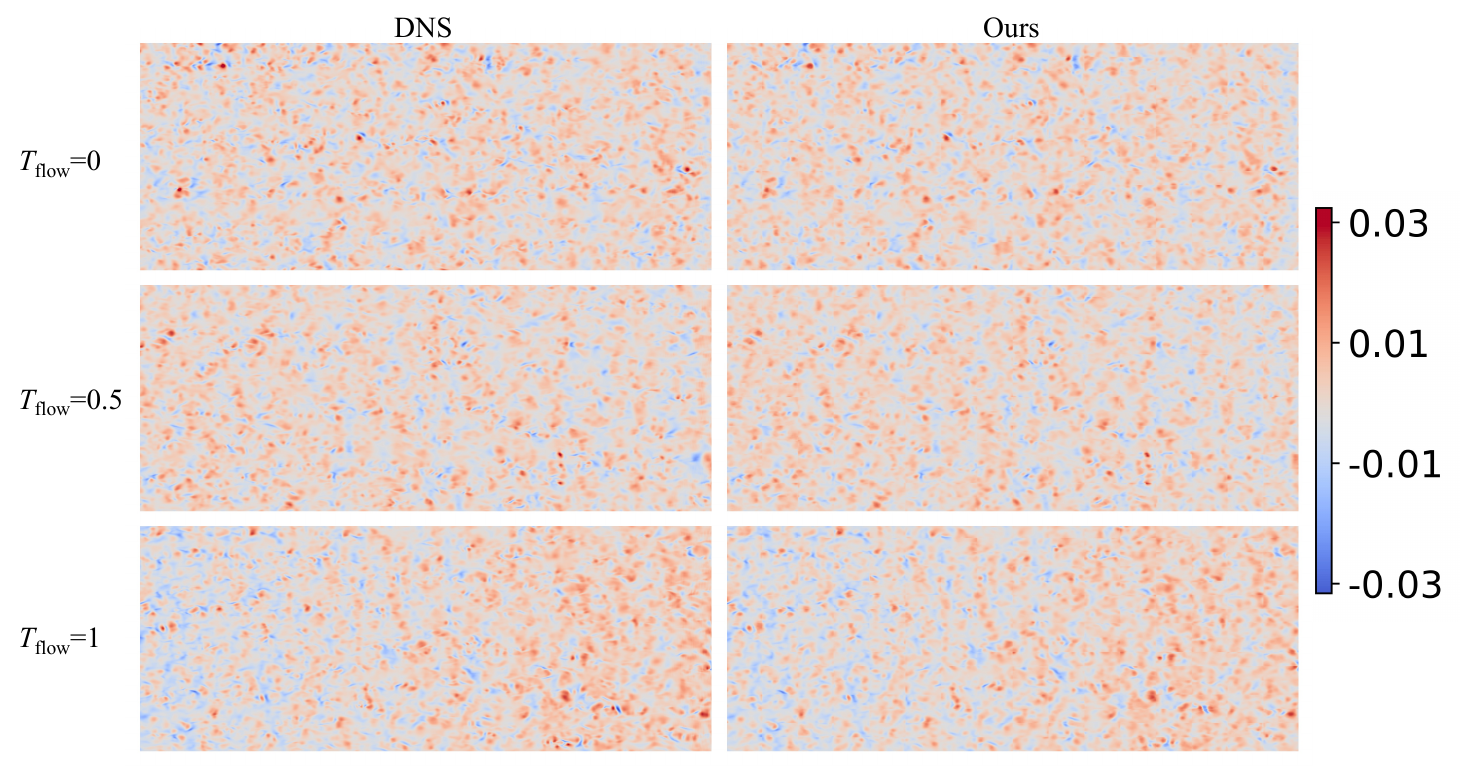}
    \caption{Reconstructed wall-pressure fluctuations for Case V0 (ZPG, $\beta = 0$), conditioned on sensor arrays at unseen time steps.}
    \label{fig:apg_v0_contour}
\end{figure}

\begin{figure}[H]
    \centering
    \includegraphics[width=0.9\textwidth]{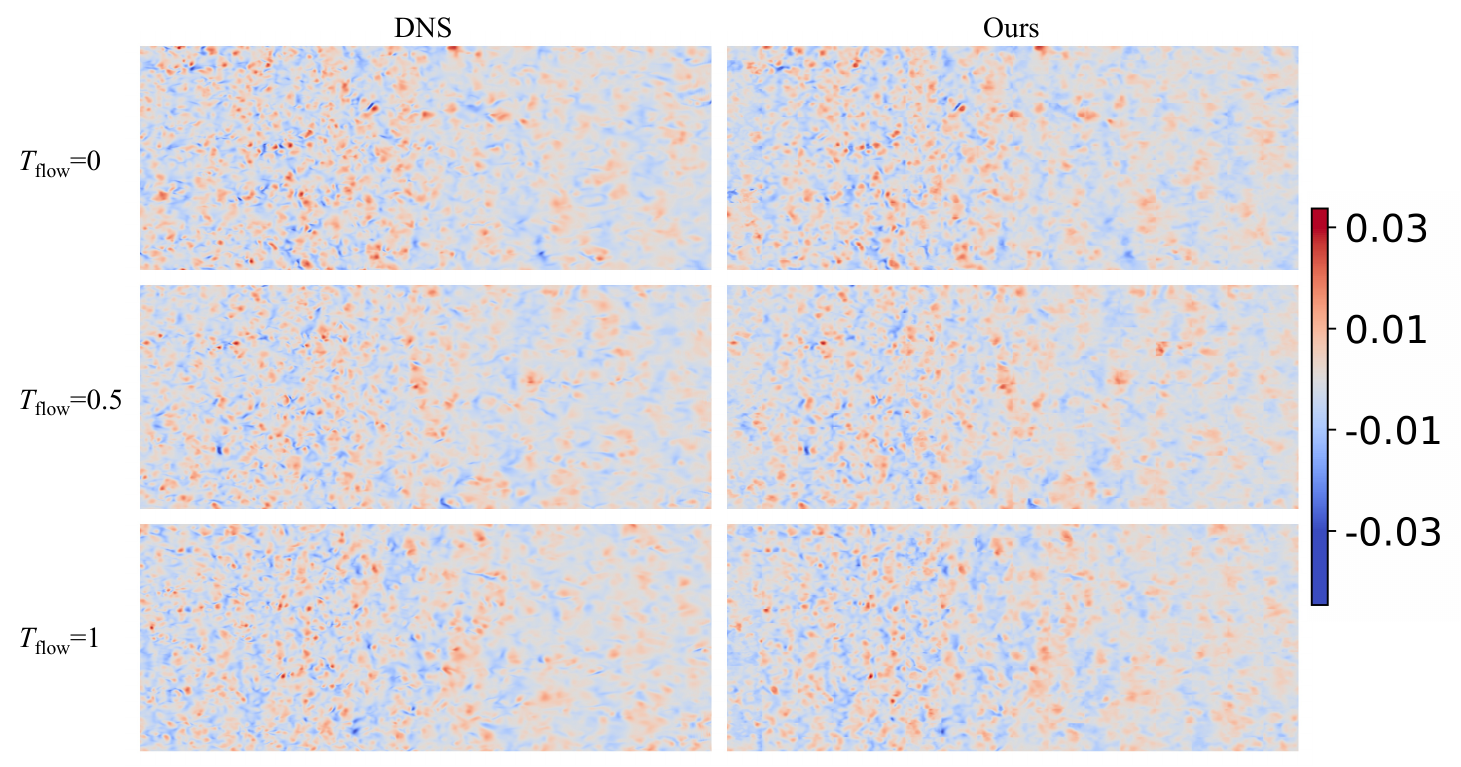}
    \caption{Reconstructed wall-pressure fluctuations for Case V40 (APG/FPG, $\beta = 6.92$), conditioned on sensor arrays at unseen time steps}
    \label{fig:apg_v40_contour}
\end{figure}

\begin{figure}[H]
    \centering
    \includegraphics[width=0.9\textwidth]{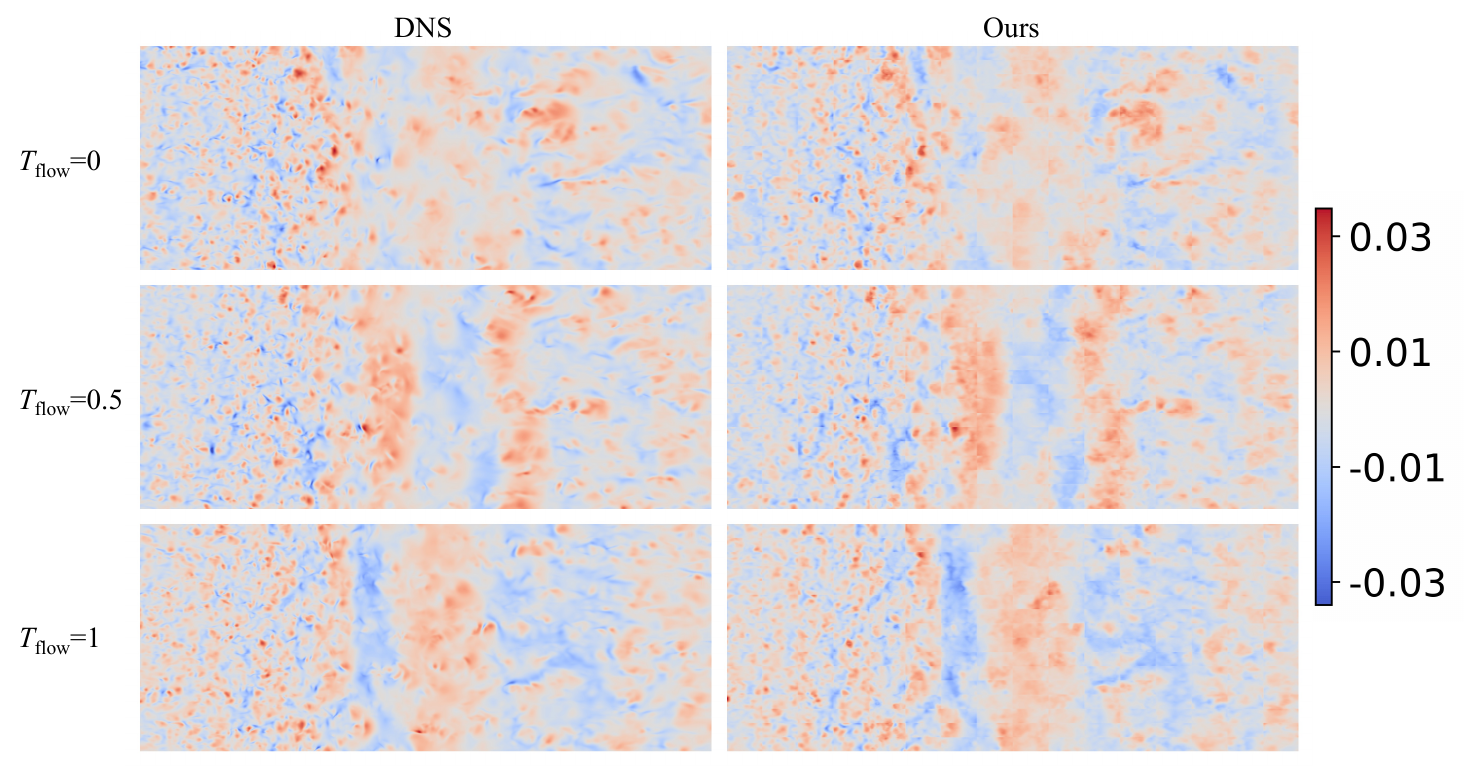}
    \caption{Reconstructed wall-pressure fluctuations for Case V80 (APG/FPG, $\beta = 62.31$), conditioned on sensor arrays at unseen time steps}
    \label{fig:apg_v80_contour}
\end{figure}

Figure~\ref{fig:apg_v0_contour} shows instantaneous reconstructions for V0 over one flow-through time. Even without an imposed pressure gradient, $p_w$ exhibits rich small-scale, convecting structure; the reconstructions closely track the DNS both frame-by-frame and across the window. As the APG/FPG strength increases to V40 ($\beta=6.92$; Fig.~\ref{fig:apg_v40_contour}), larger coherent footprints emerge downstream, marking the onset toward separation; these features are faithfully reproduced by the model. Under strong APG/FPG (V80, $\beta=62.31$; Fig.~\ref{fig:apg_v80_contour}), the DNS shows wall-pressure signatures concentrated near separation and reattachment; the reconstructions recover these large-scale footprints and their temporal evolution while maintaining the embedded small-scale fluctuations. Across all three regimes, DPS-conditioned samples remain spatiotemporally consistent with DNS, indicating that the latent diffusion prior and D-CNF decoder together can accurately reconstruct the instantaneous wall-pressure fluctuation fields.

\begin{figure}[H]
    \centering
    \includegraphics[width=0.7\textwidth]{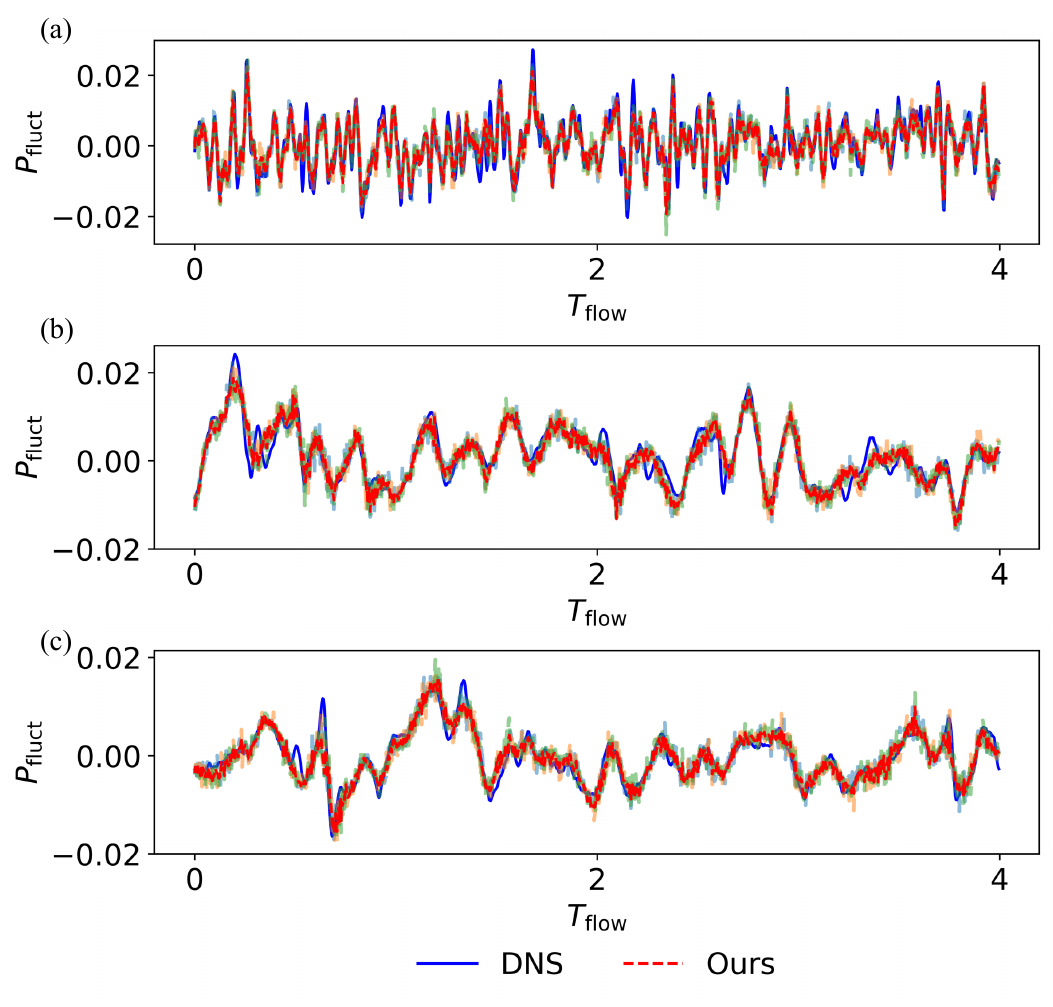}
    \caption{Time series of wall-pressure fluctuations at three unobserved locations (no sensors) for Case V80 (APG/FPG, $\beta = 62.31$), including three different sample realizations.}
    \label{fig:apg_v80_signal}
\end{figure}

At \emph{unobserved} (i.e., unsensed) locations, the reconstructions are essentially indistinguishable from DNS in both amplitude and phase; the three independent posterior samples lie tightly on the DNS trace with only a very thin scatter band (Fig.~\ref{fig:apg_v80_signal}), indicating highly accurate recovery away from sensors. Although the $16{\times}16$ array imposes strong conditioning and can reduce sample diversity, the generator still retains stochastic flexibility: small but physically consistent variations persist across posterior draws, reflecting calibrated uncertainty rather than model bias. The same behavior is observed for V0 and V40 (Figs.~\ref{fig:single_apg_v0_signal}, \ref{fig:single_apg_v40_signal}), confirming that within a fixed flow regime the method produces DNS-quality reconstructions at unsensed points while still expressing modest, plausible variability.

\begin{figure}[t!]
    \centering
    \includegraphics[width=0.9\textwidth]{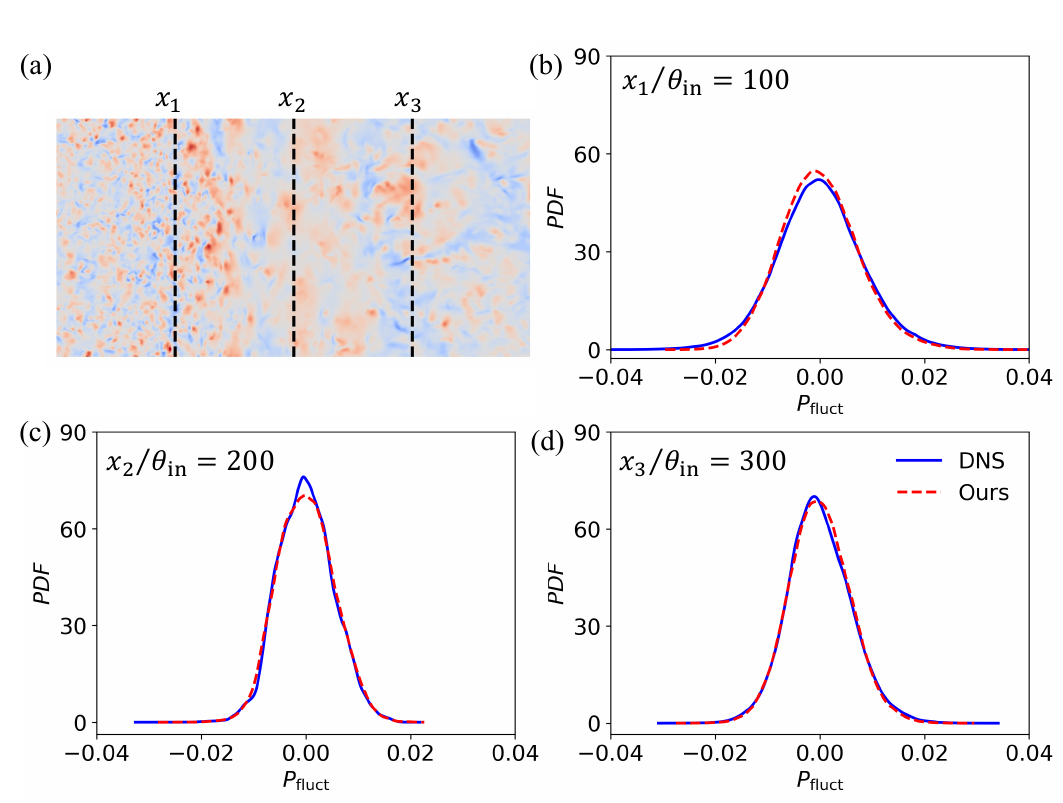}
    \caption{Probability density functions (PDFs) of wall-pressure fluctuations at three streamwise locations for Case V80 (APG/FPG, $\beta = 62.31$): (a) station locations; (b) PDF near the separation point $x_1/\theta_{\mathrm{in}} = 100$; (b) PDF near the bubble center $x_2/\theta_{\mathrm{in}} = 200$; (c) PDF near the reattachment point $x_3/\theta_{\mathrm{in}} = 300$.}
    \label{fig:apg_v80_pdf}
\end{figure}

We next assess statistical fidelity at three representative streamwise stations for V80: the separation region ($x_1/\theta_{\mathrm{in}}{=}100$), the center of the recirculation bubble ($x_2/\theta_{\mathrm{in}}{=}200$), and the reattachment region ($x_3/\theta_{\mathrm{in}}{=}300$). Figure~\ref{fig:apg_v80_pdf} compares the probability density functions (PDFs) of wall-pressure fluctuations at these locations. The reconstructed distributions closely match DNS in mean and variance and capture the tail behavior across all stations. Small discrepancies are limited to a mild underestimation of the most negative tail events near separation, while the bubble center and reattachment stations are nearly indistinguishable from DNS. Consistent agreement is observed for V0 and V40 (Figs.~\ref{fig:single_apg_v0_pdf} and \ref{fig:single_apg_v40_pdf}). 

\begin{figure}[t!]
    \centering
    \includegraphics[width=1.0\textwidth]{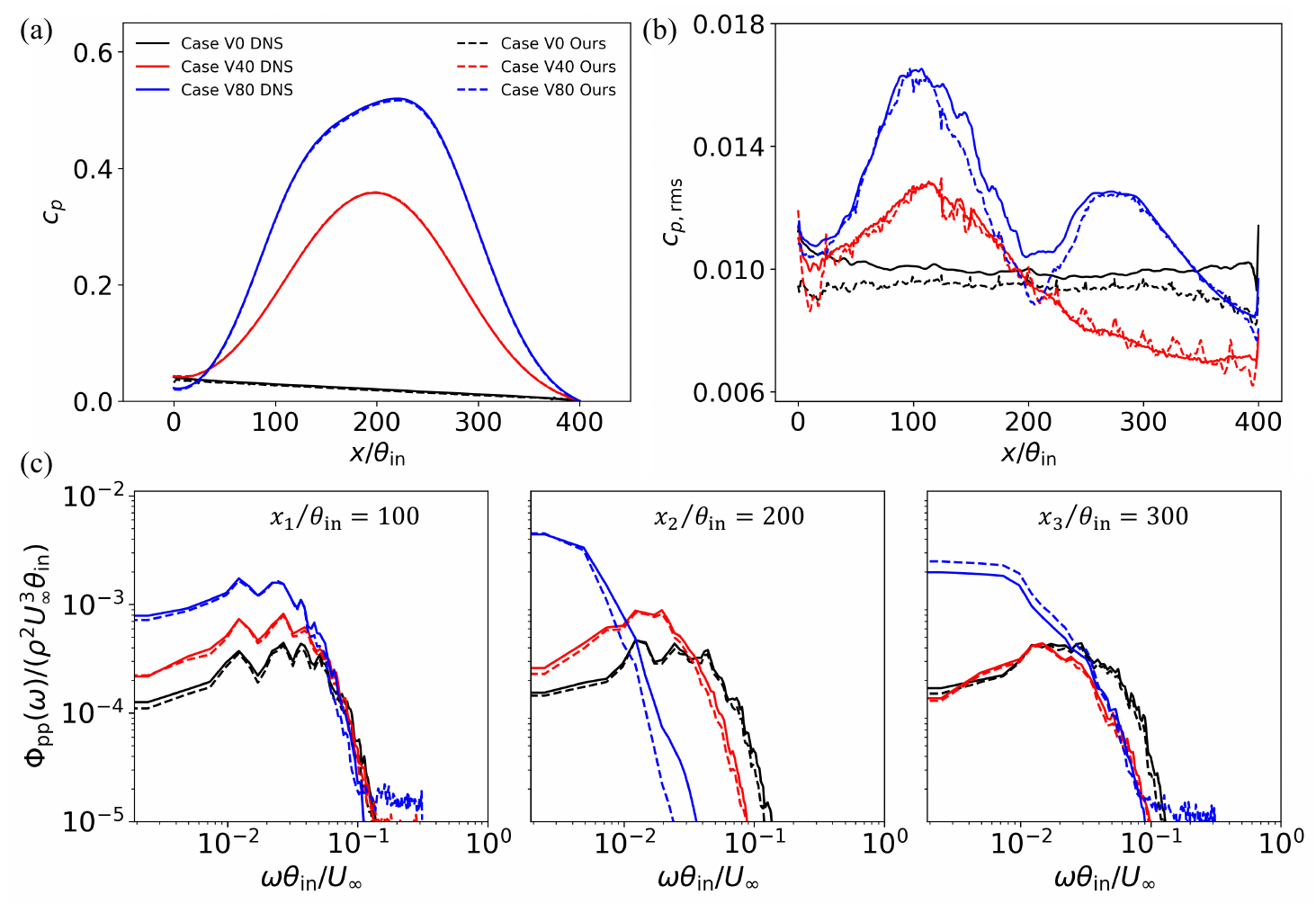}
    \caption{Turbulence statistics for all training ZPG and APG/FPG regimes on unseen time windows: (a) mean pressure coefficient; (b) root-mean-square of wall-pressure fluctuations; (c) wall-pressure frequency spectra at three representative streamwise locations.}
    \label{fig:training_stat}
\end{figure}

A more comprehensive statistical comparison across the three training pressure-gradient conditions is shown in Figure~\ref{fig:training_stat}. The mean wall-pressure coefficient $C_p$ profile is reproduced with high fidelity (panel a), including the growth and downstream decay of the APG-induced hump in V40 and V80. The root-mean-square (rms) pressure fluctuations $C_{p,\mathrm{rms}}$ are also captured reasonably well (panel b): the model recovers the single peak associated with the incipient-separation region in V40 and the two distinct peaks near separation and reattachment in V80. Frequency spectra at three streamwise stations (panel c) agree closely with DNS, preserving both the low-frequency energy associated with large scales and the correct high-frequency roll-off. Small discrepancies appear near the bubble center ($x_2/\theta_{\mathrm{in}}=200$) in V80, where the flow is most sensitive to local variations, leading to a mild mismatch around intermediate frequencies; elsewhere the agreement is excellent.

\begin{figure}[t!]
    \centering
    \includegraphics[width=1.0\textwidth]{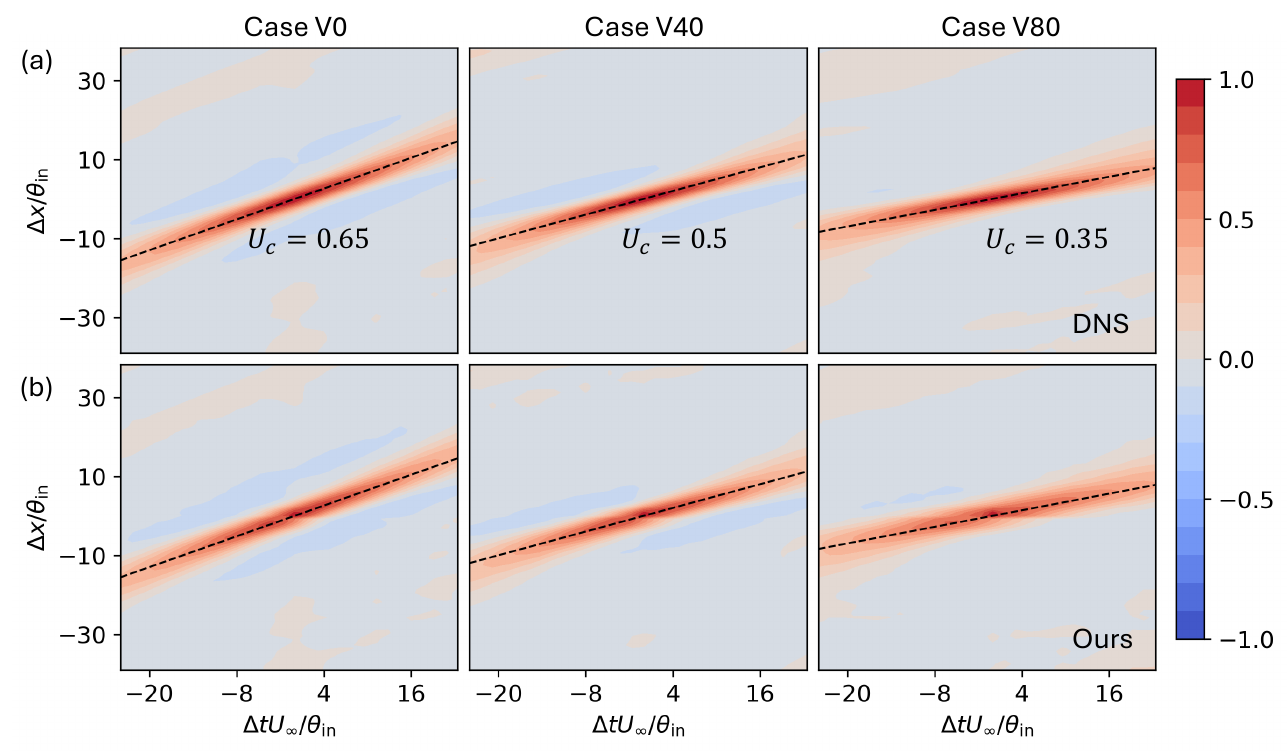}
    \caption{Contours of the streamwise space-time correlation coefficient at the streamwise location $x/ \theta_{\mathrm{in}} = 100$ for training ZPG and APG/FPG cases: (a) DNS results; (b) reconstructed field. Slopes of the dashed lines indicate the convection velocities $U_c$.}
    \label{fig:training_coe}
\end{figure}

In convection-dominated TBLs, coherent structures are organized by a local convection velocity $U_c$: at a fixed point the wall pressure signal decorrelates rapidly in time, whereas along a trajectory moving with speed $U_c$ coherence persists. This path corresponds to the ridge of maximum streamwise space-time correlation as discussed by Na and Moin~\cite{na1998direct}. To assess whether the reconstructions preserve this organization, we compute the streamwise space-time correlation coefficient $R_{pp}(\Delta x,\Delta t)$ at several stations, defined as
\begin{equation}
    R_{pp}(\Delta x,\Delta t;x) =\frac{\langle p_w(x,z,t)p_w(x+\Delta x,z,t+\Delta t)\rangle}{\sqrt{\langle p_w^2(x,z,t)\rangle}\sqrt{\langle p_w^2(x+\Delta x,z,t) \rangle}},
\end{equation}
where $\langle\cdot\rangle$ denotes time and spanwise averaging. The convection speed is estimated from the ridge slope, $U_c\simeq \Delta x/\Delta t$ along the locus of maximal correlation.
Figure~\ref{fig:training_coe} (station $x/\theta_{\mathrm{in}}=100$) shows that the reconstructions reproduce the DNS ridge angle, anisotropy, and decorrelation rates in both $\Delta x$ and $\Delta t$. As pressure-gradient strength increases (V0 $\rightarrow$ V40 $\rightarrow$ V80) the ridge flattens, indicating a decrease in $U_c$; the reconstructed fields match this trend quantitatively. 

At the bubble center $x/\theta_{\mathrm{in}} = 200$, the DNS exhibits a markedly reduced $U_c$ and longer temporal coherence for V80 due to the recirculation region. Our model recovers both the reduced ridge slope and the elongated time correlation, as shown in Fig.~\ref{fig:training_coe_x200}. Downstream at $x/\theta_{\mathrm{in}}=300$ (Fig.~\ref{fig:training_coe_x300}), where the flow reattaches to the wall, the convection velocities for all cases become nearly identical, indicating a recovery to a more uniform flow regime, again captured by the reconstruction. These observations confirm that the proposed model not only reconstructs instantaneous flow fields but also faithfully preserves the temporal/spatial coherence properties inherent to convection-dominated wall-pressure dynamics.

\subsection{Generative wall-pressure reconstruction under different pressure gradients (cross-regime testing)}
\label{sec:guidance}

We now test generalization to \emph{unseen} pressure-gradient conditions with a \emph{single} unified model. Unlike Sec.~\ref{sec:cond_gen_sensor}, where a separate model was trained per regime, here we jointly train on V0, V40, and V80 while conditioning on a continuous flow descriptor (Condition~II): the mean wall-pressure profile $C_p$ embedded via CFG. During inference, no retraining is performed: we supply the appropriate descriptor $c$ and assimilate sparse pressure sensor data (Condition~I) via DPS, enabling zero-shot reconsturction to new APG/FPG strengths. We evaluate on two \emph{interpolated} cases (V20, V60) and two \emph{extrapolated} cases (V90, V100), none used in training.

\begin{figure}[htp!]
    \centering
    \includegraphics[width=1.0\textwidth]{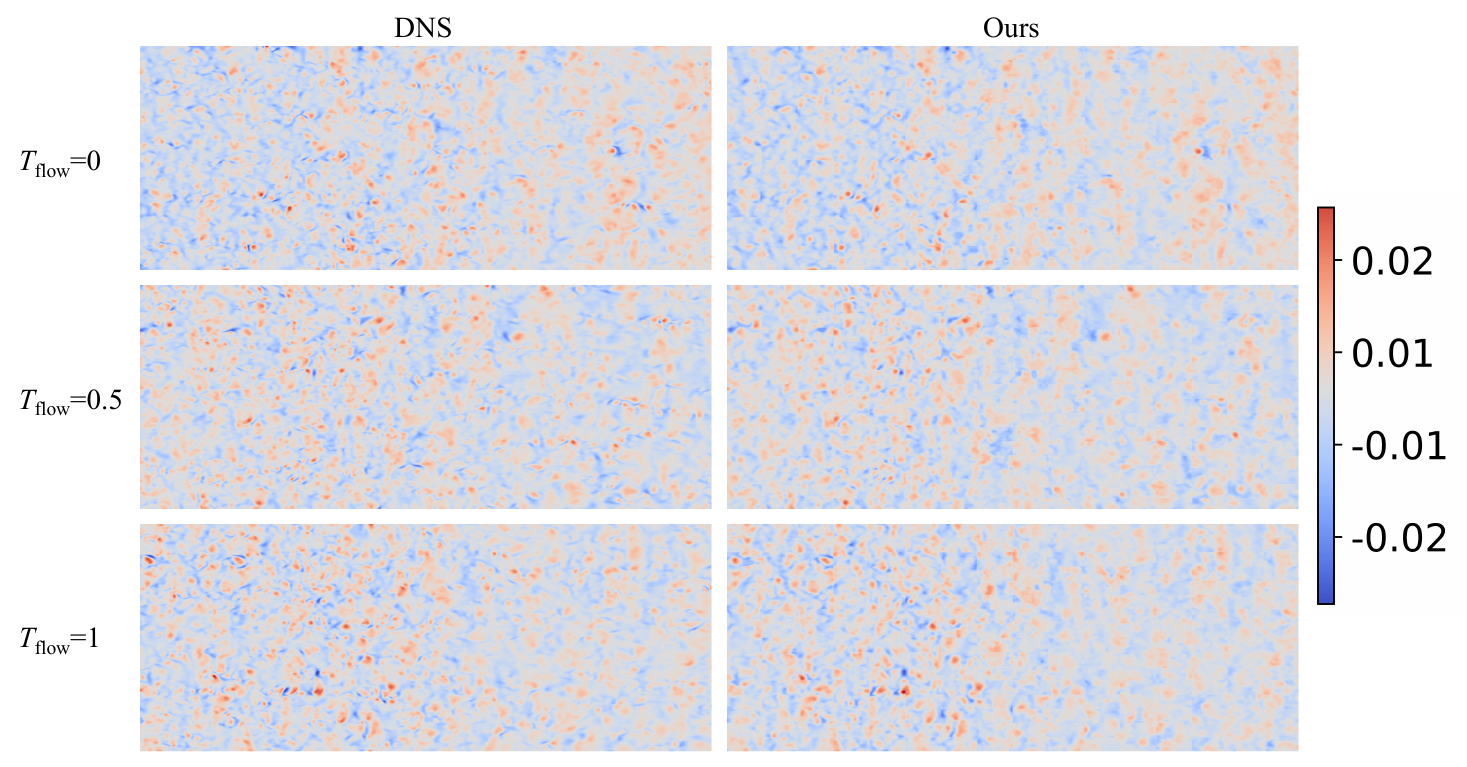}
    \caption{Reconstructed wall-pressure fluctuations for interpolated Case V20 (APG/FPG, $\beta = 2.14$), conditioned on mean profiles and sensor arrays.}
    \label{fig:apg_v20_contour}
\end{figure}

\begin{figure}[htp!]
    \centering
    \includegraphics[width=1.0\textwidth]{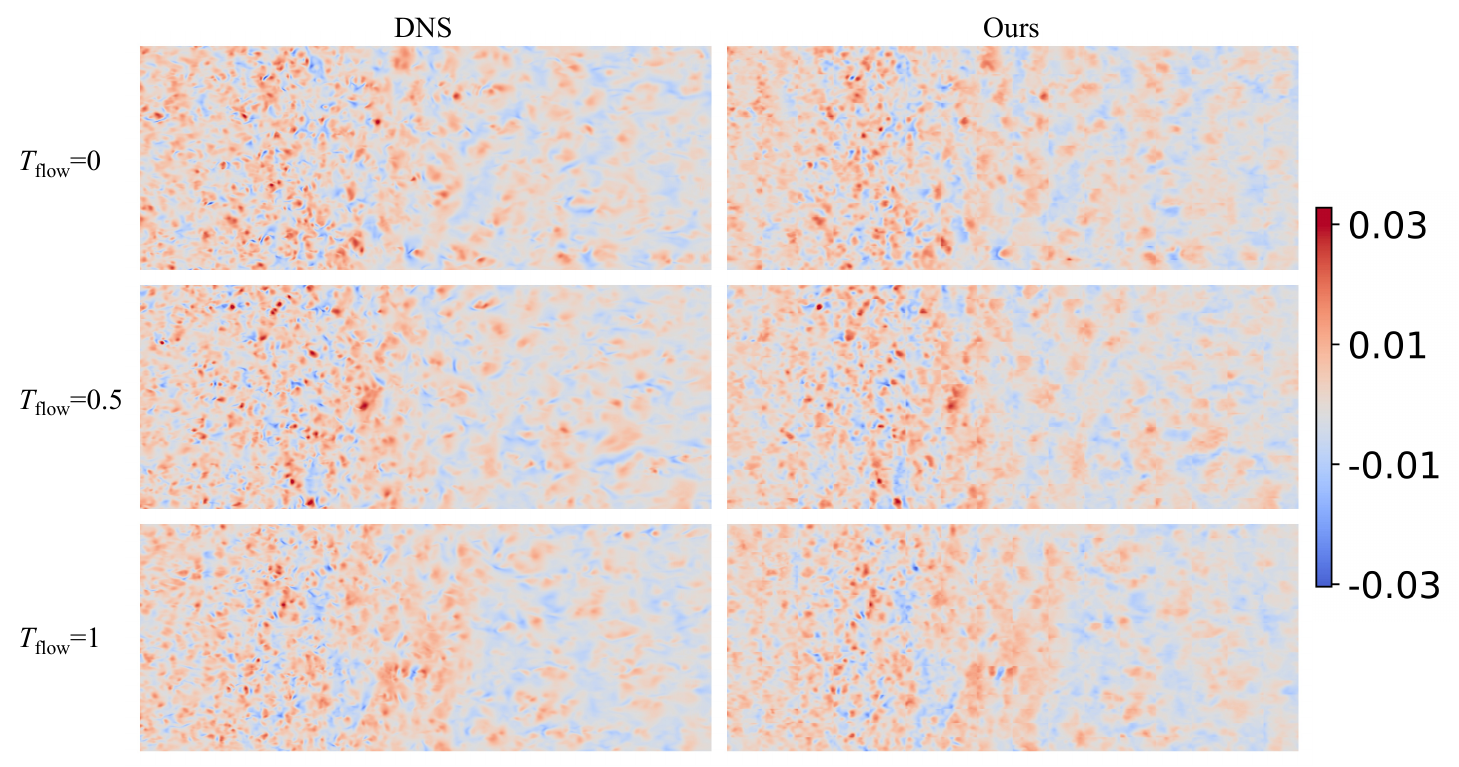}
    \caption{Reconstructed wall-pressure fluctuations for interpolated Case V60 (APG/FPG, $\beta = 22.31$), conditioned on mean profiles and sensor arrays.}
    \label{fig:apg_v60_contour}
\end{figure}

\begin{figure}[htp!]
    \centering
    \includegraphics[width=1.0\textwidth]{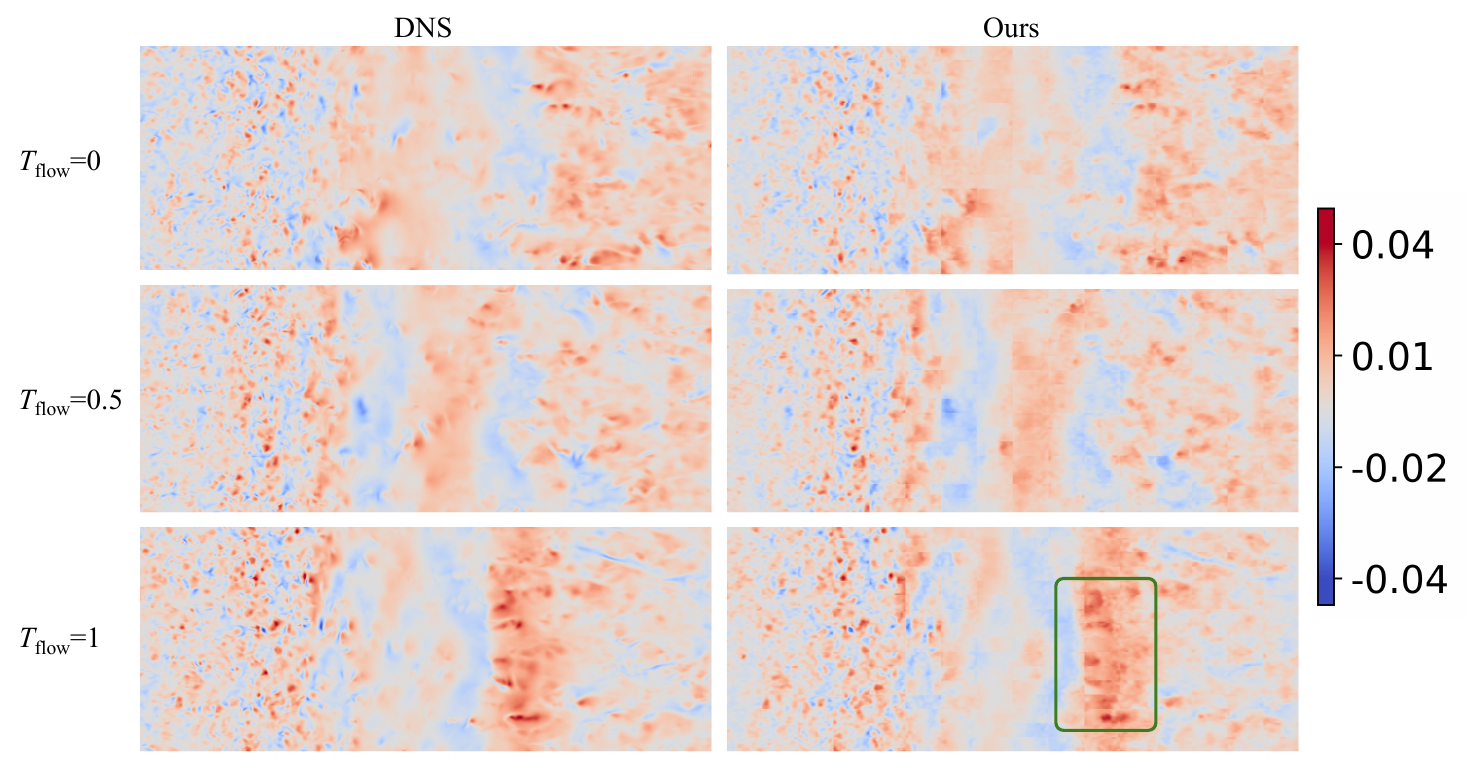}
    \caption{Reconstructed wall-pressure fluctuations for extrapolated Case V90 (APG/FPG, $\beta = 90.19$), conditioned on mean profiles and sensor arrays.}
    \label{fig:apg_v90_contour}
\end{figure}

For the interpolated cases V20 and V60, the unified model reconstructs the instantaneous wall-pressure fields with high fidelity. As shown in Fig.~\ref{fig:apg_v20_contour} and Fig.~\ref{fig:apg_v60_contour}, the instantaneous wall pressure fields are accurately reconstructed, with small-scale coherent structures effectively captured. In Case V60, where the flow remains attached to the wall, and larger structures appear in the downstream region, which are also well captured. Notably, for the extrapolated strong APG/FPG cases V90 and V100, the model successfully captures multiscale organization of separation and reattachment footprints, as illustrated in Fig.~\ref{fig:apg_v90_contour} and Fig.~\ref{fig:apg_v100_contour}; mild blurring appears near reattachment in V100, but the spatial placement and temporal evolution of large-scale structures remain accurate.

\begin{figure}[t!]
    \centering
    \includegraphics[width=0.7\textwidth]{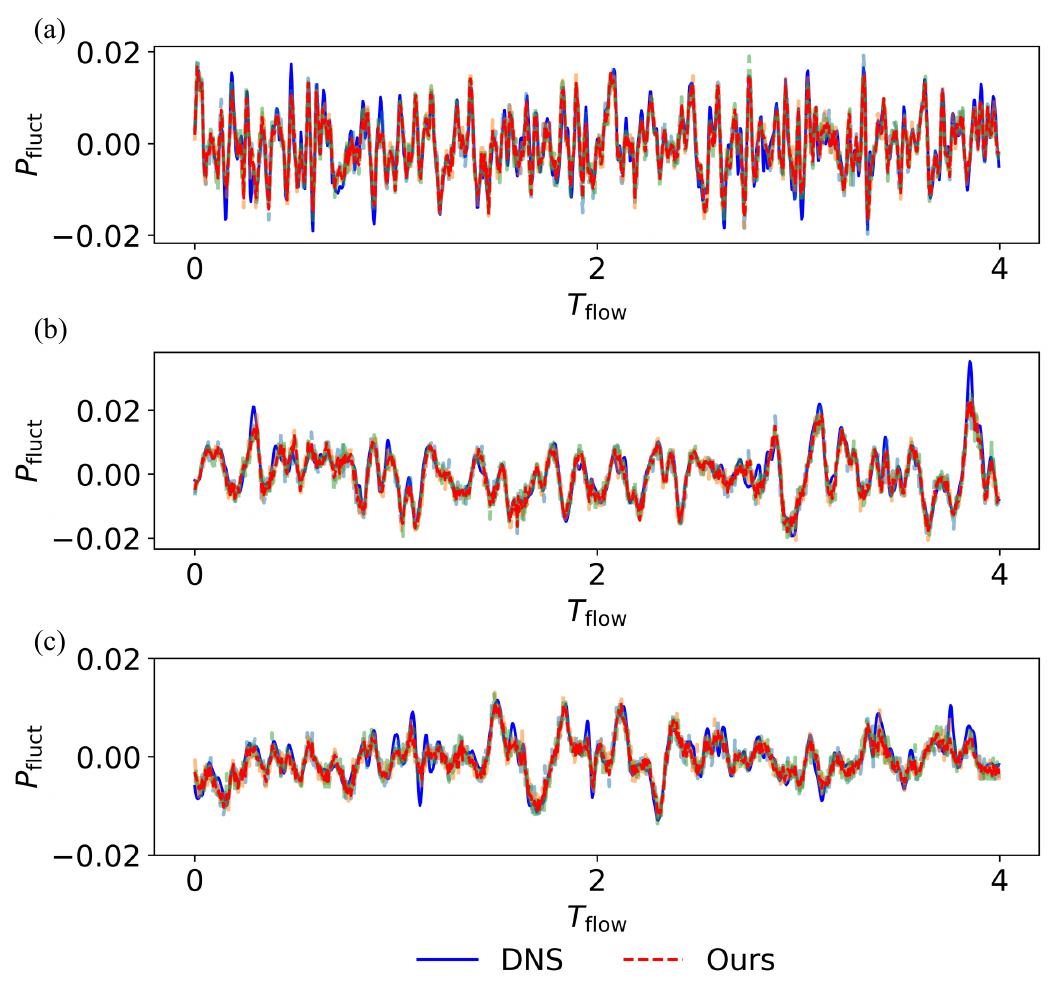}
    \caption{Time series of wall-pressure fluctuations at three unconditioned sensor locations for interpolated Case V60 (APG/FPG, $\beta = 22.31$), including three different sample realizations.}
    \label{fig:apg_v60_signal}
\end{figure}

 The generative capability of the model is further supported by the wall-pressure time series sampled at unconditioned spatial locations. As shown in Figs.~\ref{fig:apg_v60_signal}, ~\ref{fig:single_apg_v20_signal}, ~\ref{fig:single_apg_v90_signal} and ~\ref{fig:single_apg_v100_signal}, the variations across different realizations reflect the model's ability to generate diverse outputs that remain consistent with the physical characteristics of the flow.

\begin{figure}[!ht]
    \centering
    \includegraphics[width=0.9\textwidth]{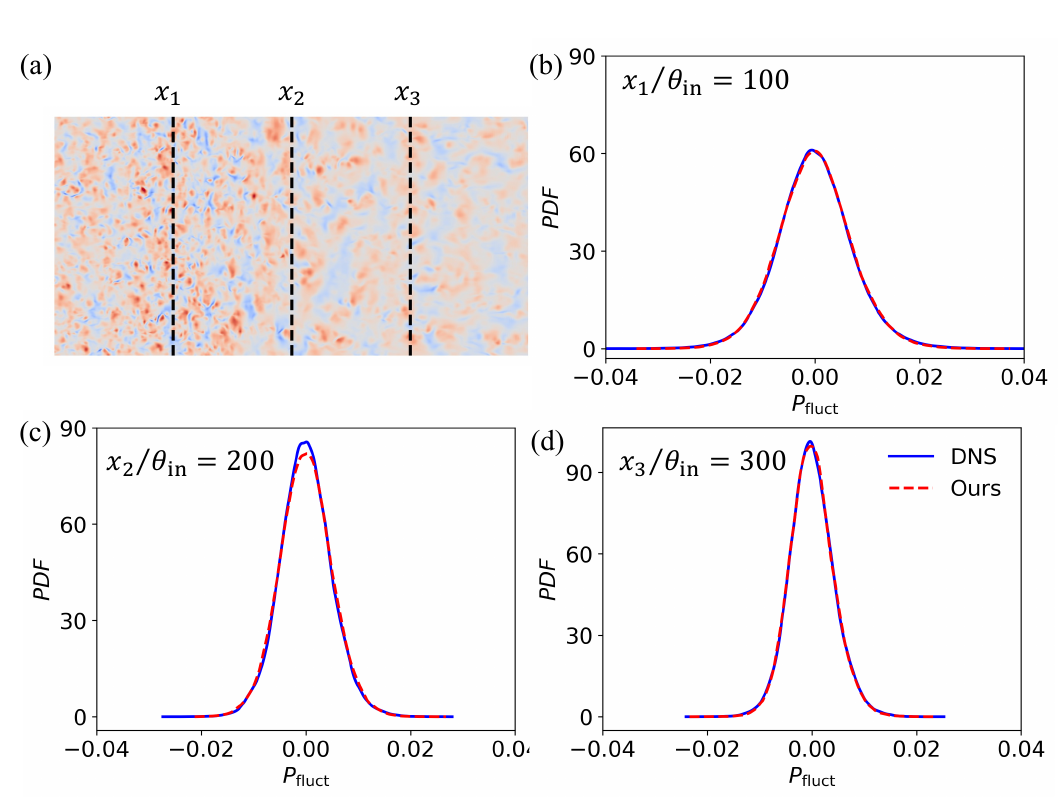}
    \caption{Probability density functions (PDFs) of wall-pressure fluctuations at three streamwise locations for interpolated Case V60 (APG/FPG, $\beta = 22.31$): (a) station locations; (b) PDF at the $x_1/\theta_{\mathrm{in}} = 100$; (b) PDF at  $x_2/\theta_{\mathrm{in}} = 200$; (c) PDF at  $x_3/\theta_{\mathrm{in}} = 300$.}
    \label{fig:apg_v60_pdf}
\end{figure}

\begin{figure}[t!]
    \centering
    \includegraphics[width=1.0\textwidth]{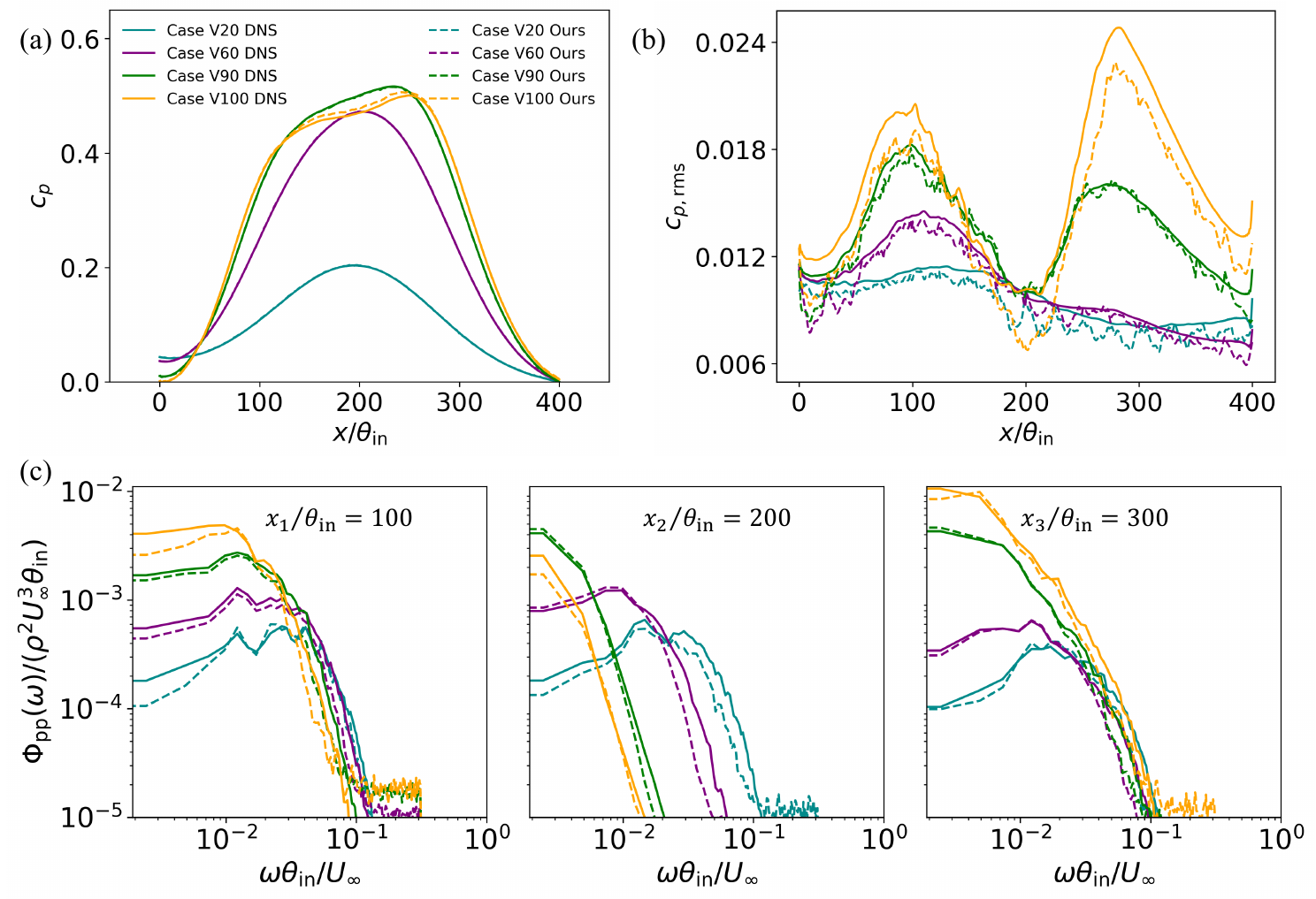}
    \caption{Turbulence statistics for all testing APG/FPG regimes: (a) mean wall pressure coefficient; (b) root-mean-square of wall-pressure fluctuations; (c) wall-presure frequency spectra at three representative streamwise locations.}
    \label{fig:single_apg_testing_stat}
\end{figure}

The reconstructed fields reproduce wall-pressure statistics with high fidelity. For V60, the PDFs of wall-pressure fluctuations at three stations (Fig.~\ref{fig:apg_v60_pdf}) closely match DNS in mean, variance, and tails; similar agreement is observed for V20, V90, and V100 (Figs.~\ref{fig:single_apg_v20_pdf}, \ref{fig:single_apg_v90_pdf}, \ref{fig:single_apg_v100_pdf}). 
Although the mean profile $C_p$ is provided at inference, it is incorporated only via unconditional sampling with probability $p_{\mathrm{uncond}}=0.2$ rather than enforced as a hard constraint. Accordingly, we recompute $C_p$ from the generated fields (Fig.~\ref{fig:single_apg_testing_stat}a): predictions are accurate for the interpolated cases (V20, V60) and remain so for the slightly extrapolated V90; a mild overprediction appears near the center of the recirculation bubble for the strongly extrapolated V100. Second-order statistics are also well captured: the peaks of root-mean-square wall-pressure fluctuations $C_{p,\mathrm{rms}}$ near separation and reattachment are generally reproduced across cases (Fig.~\ref{fig:single_apg_testing_stat}b). For V100, magnitudes are slightly underestimated, which is consistent with its being far outside the training regime. In training, the strongest APG/FPG (V80) exhibits a smaller second peak near reattachment than the first near separation (Fig.~\ref{fig:training_stat}b); V90 behaves similarly, whereas V100 shows the opposite trend (second peak exceeding the first), indicating a regime change and a modest reduction in accuracy that remains acceptable. The zigzag pattern of $C_{p,\mathrm{rms}}$ in V20 mirrors the training case V40, reflecting increased unsteadiness near separation; this local complexity occasionally introduces minor discontinuities at patch boundaries in the generated fields.
The frequency spectra are further analyzed to assess the model’s ability to reconstruct coherent structures across a broad range of spatial and temporal scales. As shown in Fig.~\ref{fig:single_apg_testing_stat}(c), the low-frequency components, associated with large-scale convective coherence, are accurately recovered. This behavior will be examined in greater detail through the subsequent analysis of space-time correlations. At higher frequencies, the reconstructed wall-pressure fields capture the majority of the small-scale fluctuations, although a slight overestimation is observed in regions corresponding to extremely low energy content. Should enhanced accuracy at small scales be required, the spectrally decomposed framework previously proposed by Fan et al.~\cite{fan2025neural} may provide an effective strategy for improving model fidelity at finer scales.

\begin{figure}[!ht]
    \centering
    \includegraphics[width=1.0\textwidth]{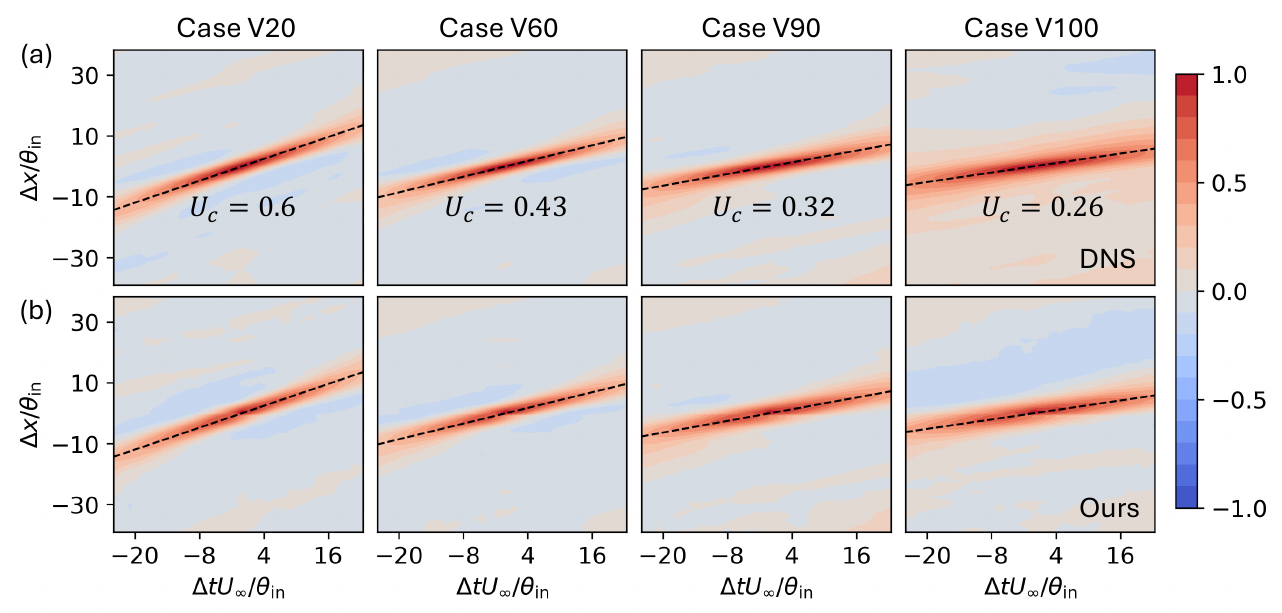}
    \caption{Contours of the streamwise space-time correlation coefficient at the streamwise location $x/ \theta_{\mathrm{in}} = 100$ for unseen APG/FPG cases: (a) DNS results; (b) reconstructed field. Slopes of the dashed lines indicate the convection velocities $U_c$.}
    \label{fig:testing_coe}
\end{figure}
Following the analysis of the frequency spectra, the space-time correlation coefficients for all testing cases are presented in Fig.~\ref{fig:testing_coe}. The convection velocities $U_c$ inferred from the reconstructed wall-pressure fields show strong agreement with those obtained from DNS. Near the separation point $x/\theta_{\mathrm{in}} = 100$, the model accurately captures the decreasing trend of
$U_c$ with increasing APG/FPG, reflecting the expected deceleration of coherent structures in the presence of stronger pressure gradients. Further validation is provided by the results near the bubble center $x/\theta_{\mathrm{in}} = 200$ (Fig.~\ref{fig:testing_coe_x200}) and the reattachment point $x/\theta_{\mathrm{in}} = 300$ (Fig.~\ref{fig:testing_coe_x300}), where the predicted convection velocities remain in close agreement with DNS. These results confirm that the model not only reconstructs the instantaneous flow structures but also faithfully preserves the underlying convective transport dynamics, which are essential for accurately representing convection dominated wall-pressure fields in turbulent flows. The quantitative and qualitative evaluations presented above collectively demonstrate the generalization capability of the proposed model in reconstructing wall-pressure fields across a range of APG/FPG conditions.

\subsection{Computational cost (forward vs.\ inverse)}

\begin{figure}[t!]
    \centering
    \includegraphics[width=0.65\textwidth]{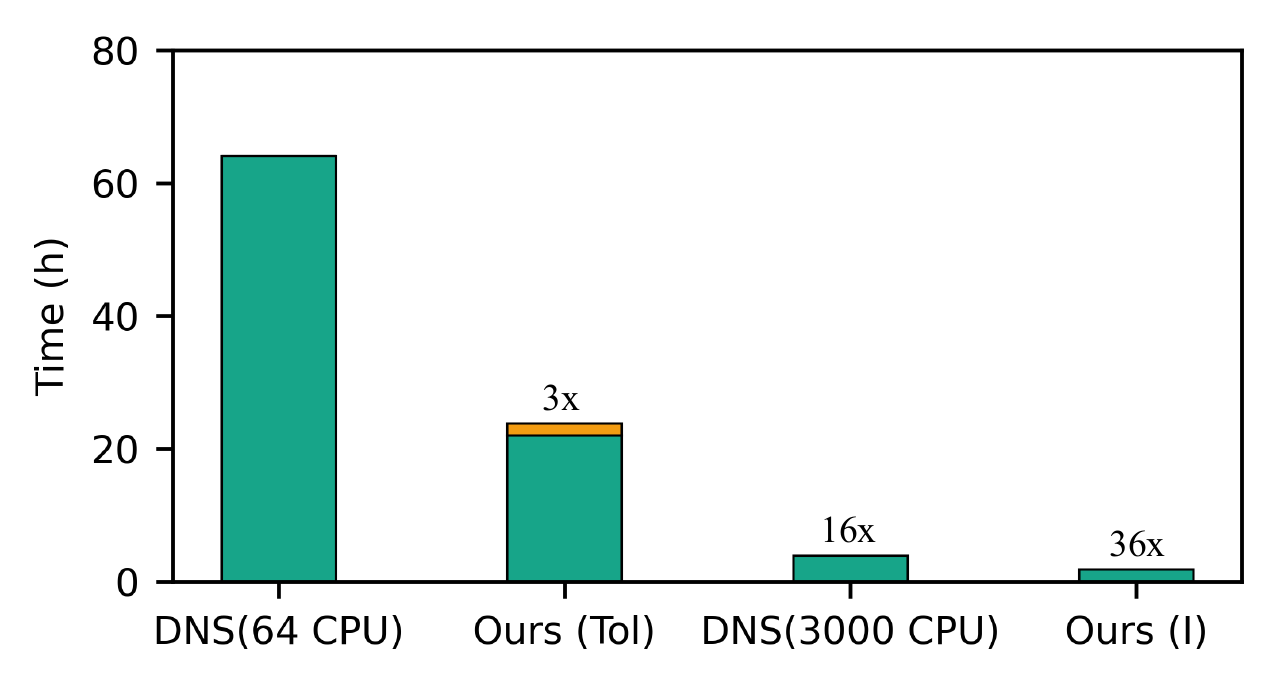}
    \caption{Wall-clock time to produce one flow-through of wall-pressure fluctuations. ``Ours (Tot)’’ includes training plus one inference; ``Ours (I)’’ is inference only. DNS bars reflect \emph{forward} runs on 64 and 3000 CPUs. For DNS-based \emph{inverse} reconstruction (e.g., adjoint or Kalman inversion), the total cost scales as multiples of the forward cost (typically $\mathcal{O}(\text{tens})\,T_{\text{fwd}}$) and is not plotted for clarity.}
    \label{fig:cost}
\end{figure}

Figure~\ref{fig:cost} reports wall-clock time to produce one flow-through of wall-pressure fluctuations. DNS timings are \textbf{forward} runs on 64 and 3000 CPUs. Counting training plus one inference (``Ours (Tot)’’), our approach is about 3$\times$ faster than DNS on 64 CPUs; inference alone (``Ours (I)’’) is about 36$\times$ faster than the 64-CPU DNS and still roughly 2$\times$ faster than DNS on 3000 CPUs. Because training is amortized across many reconstructions (different initial conditions, APGs, or long realizations), the effective speedup increases with reuse.

Crucially, DNS-based \emph{pressure-field reconstruction} is an \textbf{inverse} problem. With variational adjoint~\cite{ashtari2023identifying,wang2025variational} or Kalman-type inversion~\cite{wang2016data,wang2016quantification}, the cost scales with the number of forward/adjoint evaluations,
\begin{equation}
   T_{\text{inv}}^{\text{DNS}} \approx 
    \begin{cases}
    (N_{\text{iter}}{+}1)\,T_{\text{fwd}} & \text{(adjoint/4D-Var)}\\
    N_{\text{ens}}\,T_{\text{fwd}} & \text{(EnKF/EnKS)}
    \end{cases} 
\end{equation}
which is typically on the order of \emph{tens of forward solves} for practical windows. Thus a DNS-based inverse reconstruction would incur an order-of-magnitude (or more) higher cost than the forward bars shown, whereas our method's reported cost \emph{already} corresponds to solving the inverse task directly (conditioning on $C_p$ and generating $p_w(x,t)$). This distinction makes the practical computational advantage of the proposed approach substantially larger than suggested by forward-only comparisons.

\section{Discussion}
\label{sec:dis}
\subsection{Generalizability of the domain-decomposed CNF}
\label{sec: discuss_cnf}

Our framework relies on a {domain-decomposed conditional neural field} (D-CNF) to map each instantaneous wall-pressure field to a compact latent and decode latents back to the physical domain. Because the diffusion model operates entirely in latent space, the D-CNF with a shared SIREN backbone and fully projected hypernetwork must generalize across flow regimes and times. We therefore evaluate the D-CNF as a standalone representation model.

As described in Sec.~\ref{sec:dataset}, the training dataset covers a time window from 0 to 4 flow-through times, corresponding to 1000 snapshots. All cases share the same container parameters $(\zeta^\ast,\gamma^\ast)$, while per-tile, per-snapshot latents are optimized during training (no amortized encoder). Reconstruction accuracy is measured by the relative $L_2$ error for case $k$,
\begin{equation}
  r^k=\frac{\lVert \hat{p}_w^k (\mathbf{s}^*, \zeta^*, \gamma^*)-p_w^k \rVert _{2}}{\lVert p_w^k\rVert _{2}}
  \label{eq:relative error}
\end{equation}

\begin{figure}[htp!]
    \centering
    \includegraphics[width=1.0\textwidth]{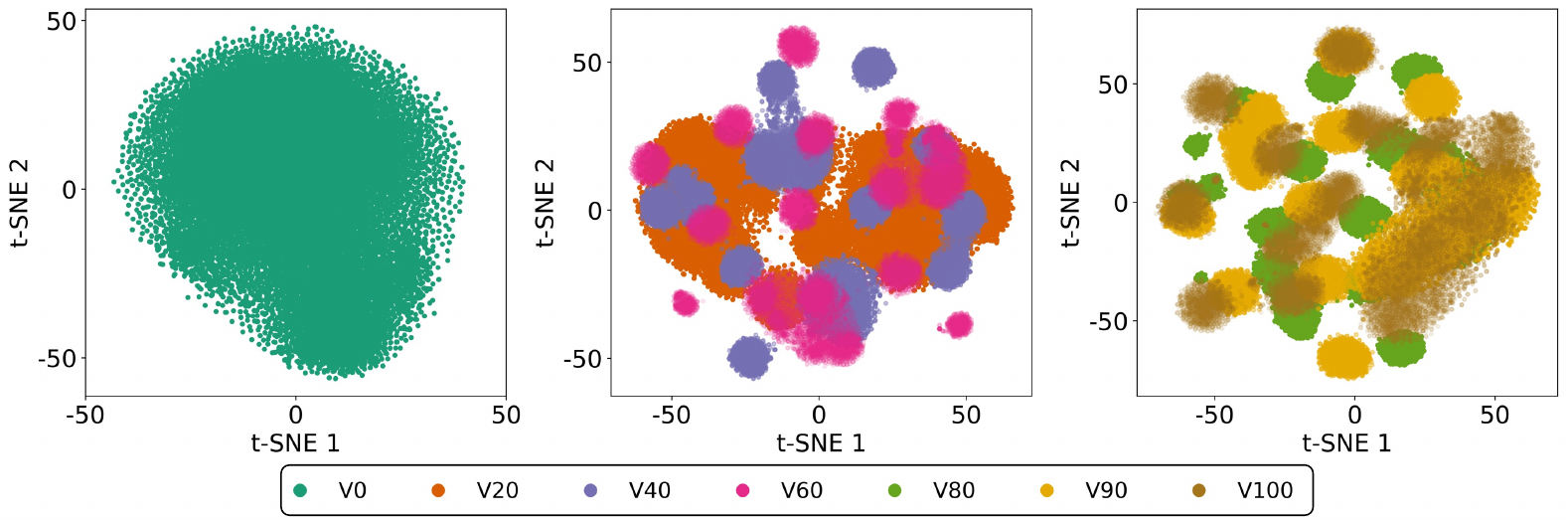}
    \caption{T-SNE analysis of the compressed latent representations from the D-CNF for training snapshots ($0-4$ flow through times): (a) zero-pressure-gradient Case V0; (b) moderate pressure-gradient Case V20, V40 and V60; and (c) strong pressure-gradient Case V80, V90 and V100.}
    \label{fig:training_latent_tsne}
\end{figure}

Table~\ref{tab:CNF_error} shows errors of $1\%$–$4\%$ for all APG/FPG cases both inside (0–4 $T_{\mathrm{flow}}$) and outside (8–12 $T_{\mathrm{flow}}$) the training window, indicating good temporal generalization once the container is fixed. The larger error for V0 ($\approx16\%$) is consistent with its nearly homogeneous, small-scale-dominated fluctuations, which offer fewer coherent patterns for the latent model to exploit. This interpretation is supported by the t-distributed stochastic neighbor embedding (t-SNE) visualizations in Fig.~\ref{fig:training_latent_tsne}: V0 yields a single diffuse cluster, whereas increasing APG/FPG produces well-separated clusters (two at V20 and multiple beyond V40), mirroring the emergence of larger coherent wall-pressure structures in the physical domain.
\begin{table}[t!]
    \centering    
    \caption{Relative $L^2$ error of the D-CNF (\%)}
    \footnotesize
    \begin{tabular}{c c c c c c c c}
    \toprule[1.5pt]
        Case & V0 & V20 & V40 & V60 & V80 & V90 & V100 \\
        \midrule
        Training (0 - 4 $T_{\mathrm{flow}}$)   &15.97 & 3.43 & 1.78 & 1.24 & 1.45 & 1.11 & 1.19\\
        Testing (8 - 12 $T_{\mathrm{flow}}$) & 15.70 & 3.45 & 2.24 & 1.24 & 1.15 & 1.07 &  1.15 \\
    \bottomrule[1.5pt] 
    \end{tabular}
    \label{tab:CNF_error}
\end{table}

To assess generalization capability, we freeze the base SIREN network parameters $(\zeta^\ast,\gamma^\ast)$ and optimize the latents (i.e., autodecoding) for completely unseen snapshots from 8-12 $T_{\mathrm{flow}}$, which are fully decorrelated from the training data. The resulting errors remain comparable to those in the training window (Tab.~\ref{tab:CNF_error}), and the reconstructed fields preserve global patterns and statistics (Fig.~\ref{fig:testing_cnf}). These results indicate that the D-CNF with non-overlapping tiling provides a robust, reusable container for convection-dominated wall-pressure fields. Empirically, a monolithic CNF without domain decomposition does not attain similar generalization on these cases, underscoring the importance of the patching strategy~\cite{guo2025conditional}.

\subsection{Analysis of diffusion-generated latents}
\label{sec:gen_latent}

\begin{figure}[htp!]
    \centering
    \includegraphics[width=1.0\textwidth]{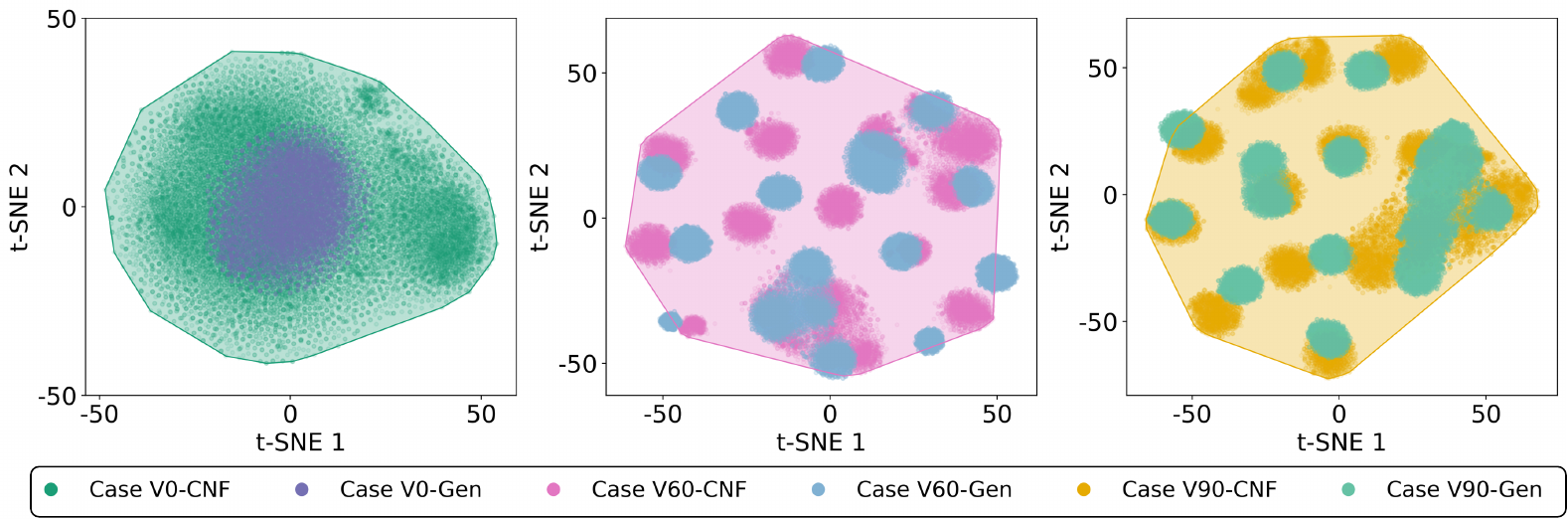}
    \caption{Comparison of the generated latent vectors and D-CNF latent representations in the T-SNE analysis for testing snapshots ($8-12$ flow through times): (a) Case V0; (b) Case V60; and (c) Case V90.}
    \label{fig:gen_latent_tsne}
\end{figure}

We examine the quality of the diffusion prior by comparing its generated latents (conditioned by sensors and/or the mean-profile descriptor) with the reference latents from the DNS snapshots (8–12 flow-through times) projected by the D-CNF. To visualize the data structure, we apply t-SNE to the combined sets for three representative regimes: the zero-pressure-gradient case V0, an interpolated case V60, and an extrapolated case V90 (Fig.~\ref{fig:gen_latent_tsne}; reference D-CNF latents are outlined).

Across all three regimes, the generated latents lie largely \emph{within the support} of the reference distribution, while remaining distinct from the exact D-CNF latents. For V0, the diffusion samples reproduce the unimodal, diffuse cloud without mode collapse. For V60, the multimodal cluster pattern is recovered and populated with comparable density, with only small cluster-wise offsets, consistent with posterior sampling under DPS and sensor noise. For V90 (extrapolated), the cluster layout is preserved with a mild covariate shift, yet samples remain inside the envelope of reference latents. Taken together, these plots indicate that the diffusion model samples new realizations on the learned D-CNF manifold (no memorization) while staying close enough to enable faithful decoding back to the physical space. We emphasize that t-SNE is used for qualitative assessment of support and mode coverage; distances are not interpreted quantitatively.

\subsection{Novelty, limitation and future prospects}

Reconstructing flow fields from sparse sensors or low-resolution data has been studied for decades, with most methods being deterministic and focused on velocity. In contrast, our framework is probabilistic: a latent diffusion model, coupled with a D-CNF, reconstructs the full spatiotemporal wall-pressure field with calibrated uncertainty. We demonstrate two primary capabilities: (i) generating new flow realizations under fixed parameters and (ii) zero-/few-shot generalization across APG/FPG regimes via classifier-free guidance and DPS assimilation. As a result, the method serves as an efficient surrogate able to produce many physically plausible wall-pressure fields conditioned on available sensors.

The main limitation stems from the local factorization used for efficiency. To handle high-dimensional, convection-dominated structure, the D-CNF compresses the domain into non-overlapping patches and the diffusion prior operates on per-patch latents. While effective, this design weakens cross-patch coupling: without sufficient measurements, reconstructions can be realistic within each patch yet exhibit misalignment across patches. In the current implementation we therefore use a relatively dense $16{\times}16$ sensor array to anchor each patch. Future work will focus on enforcing global spatial coherence with fewer sensors, for example, overlapping tiles with interface penalties on $p_w$ and $\nabla p_w$, spatial autoregression or attention that exchanges boundary context between neighboring patches, and hierarchical latents that combine a coarse global code with fine local codes.

A second limitation is computational. Compared with our earlier global reconstruction variants~\cite{du2024conditional, liu2024confild}, patch-based diffusion has higher inference cost, dominated by denoising diffusion implicit model (DDIM) sampling (typically $\ge 300$ steps for accurate latents). Promising accelerations include single/few-step diffusion~\cite{yin2024one} or consistency models~\cite{song2023consistency}, progressive/distillation-style samplers, and improved latent compression (e.g., coarser global latents with patch refinements). These directions should reduce runtime while preserving fidelity and coherence at the domain scale.

\section{Conclusions}
\label{sec:conclusion}
Reconstructing fully resolved, spatiotemporal wall-pressure fields in turbulent boundary layers is a challenging, ill-posed inverse problem. We presented a probabilistic generative framework that infers complete wall-pressure fields from sparse sensors, rather than predicting spectra alone. The method couples a \emph{patchwise, domain-decomposed conditional neural field} (D-CNF) for compact, high-resolution representations with a \emph{latent diffusion} prior that samples physically plausible realizations.
Two complementary conditioning mechanisms enable generalization and training-free data assimilation. (i) \emph{Diffusion posterior sampling} (DPS) incorporates arbitrary sensor layouts \emph{only at inference}, decoupling training from any specific instrumentation. (ii) \emph{Classifier-free regime guidance} integrates a low-cost flow descriptor during training and inference, promoting zero-/few-shot transfer across different APG/FPG conditions. 

Across TBL cases spanning weak to strong pressure gradients, the framework reconstructs instantaneous fields with high fidelity and recovers key statistics, including PDFs, $C_p$, $C_{p,\mathrm{rms}}$, frequency spectra, space-time correlation, and convection velocity, while providing calibrated uncertainty via ensembles. These results indicate that the approach functions as an efficient, generalizable surrogate for wall-pressure dynamics under sparse sensing.

Limitations point to clear next steps. The patchwise factorization can introduce cross-patch seams under very sparse sensing; enforcing continuity via overlapping tiles with interface penalties, boundary-aware attention/autoregression, or hierarchical (global-local) latents is a priority. Inference cost is dominated by diffusion sampling; adopting distilled or consistency-based few-/single-step solvers, together with improved latent compression, should further reduce runtime. Extensions to more complex geometries, higher Reynolds numbers, and multi-physics couplings are natural directions for future work.

\section*{Acknowledgements}
 The authors would like to acknowledge the funds from Office of Naval Research under award numbers N00014-23-1-2071 and National Science Foundation under award numbers OAC-2047127. 
 
\section*{Compliance with Ethical Standards}
Conflict of Interest: The authors declare that they have no conflict of interest.

\appendix
\section{Architecture and parameters of domain-decomposed CNF}
\label{sec: arch_para_pcnf}
\begin{figure}[htp!]
    \centering
    \includegraphics[width=0.7\textwidth]{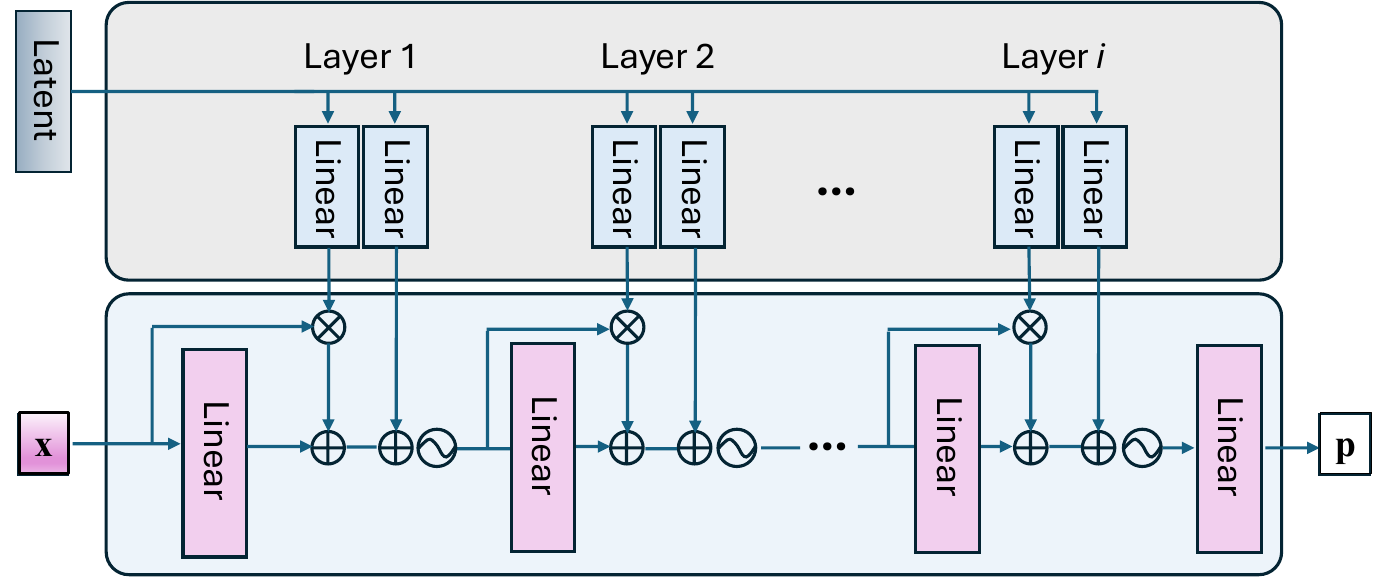}
    \caption{Architecture of the domain-decomposed CNF. The top section represents the conditional components, while the bottom section depicts the container.}
    \label{fig:schematic_pcnf}
\end{figure}

The overall architecture of the D-CNF used in this study is illustrated in Fig.~\ref{fig:schematic_pcnf}. The latent dimension is set to 64. Both the container and conditional components consist of six layers, with each layer comprising 64 neurons.

\section{Architecture and parameters of latent diffusion model}
\label{sec: arch_para_diffusion}

\begin{figure}[H]
    \centering
    \includegraphics[width=0.85\textwidth]{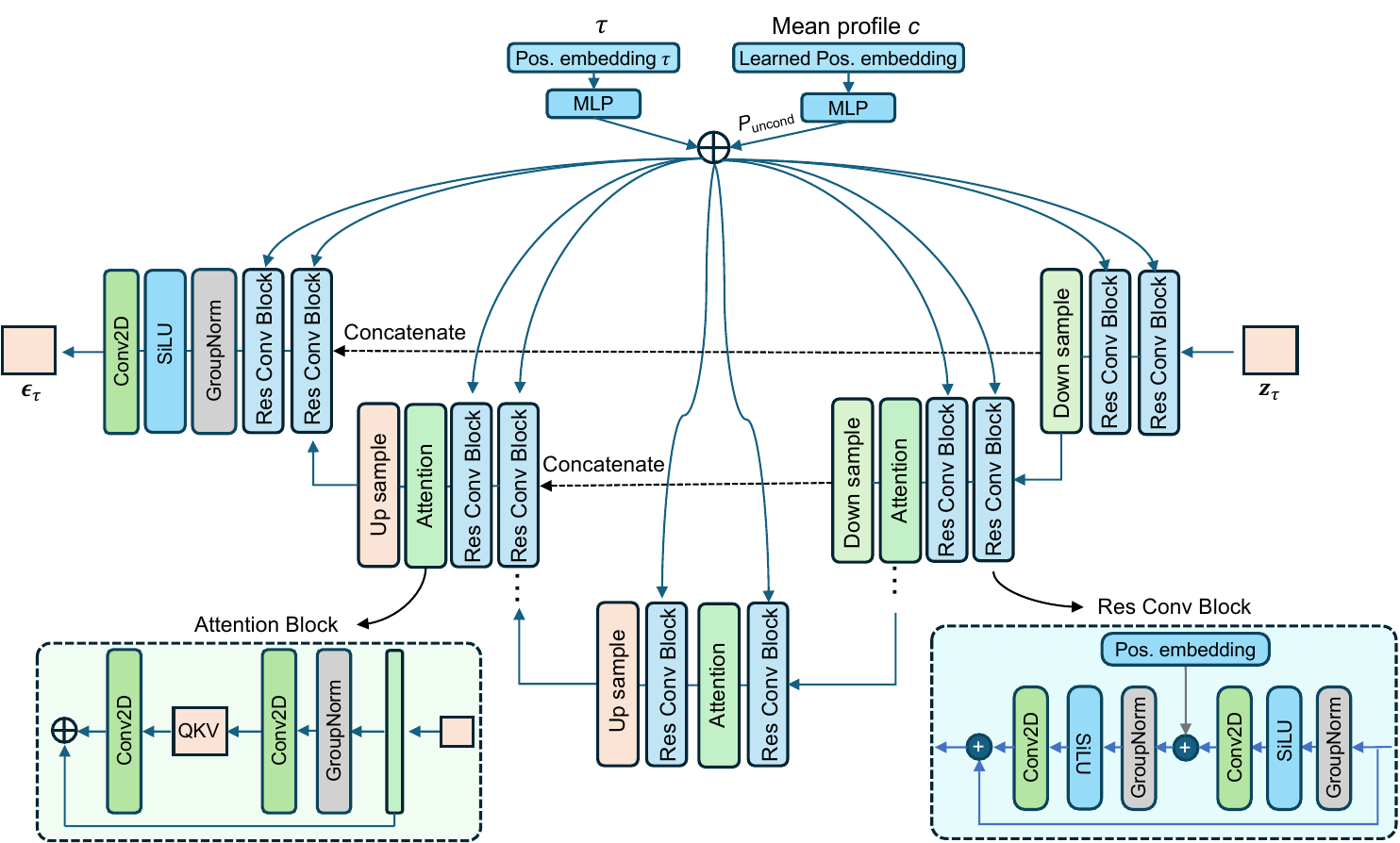}
    \caption{U-Net architecture employed as the denoising component within the latent diffusion model.}
    \label{fig:schematic_diffusion}
\end{figure}

\section{Additional results}
\label{sec: additional_results}

\begin{figure}[H]
    \centering
    \includegraphics[width=1.0\textwidth]{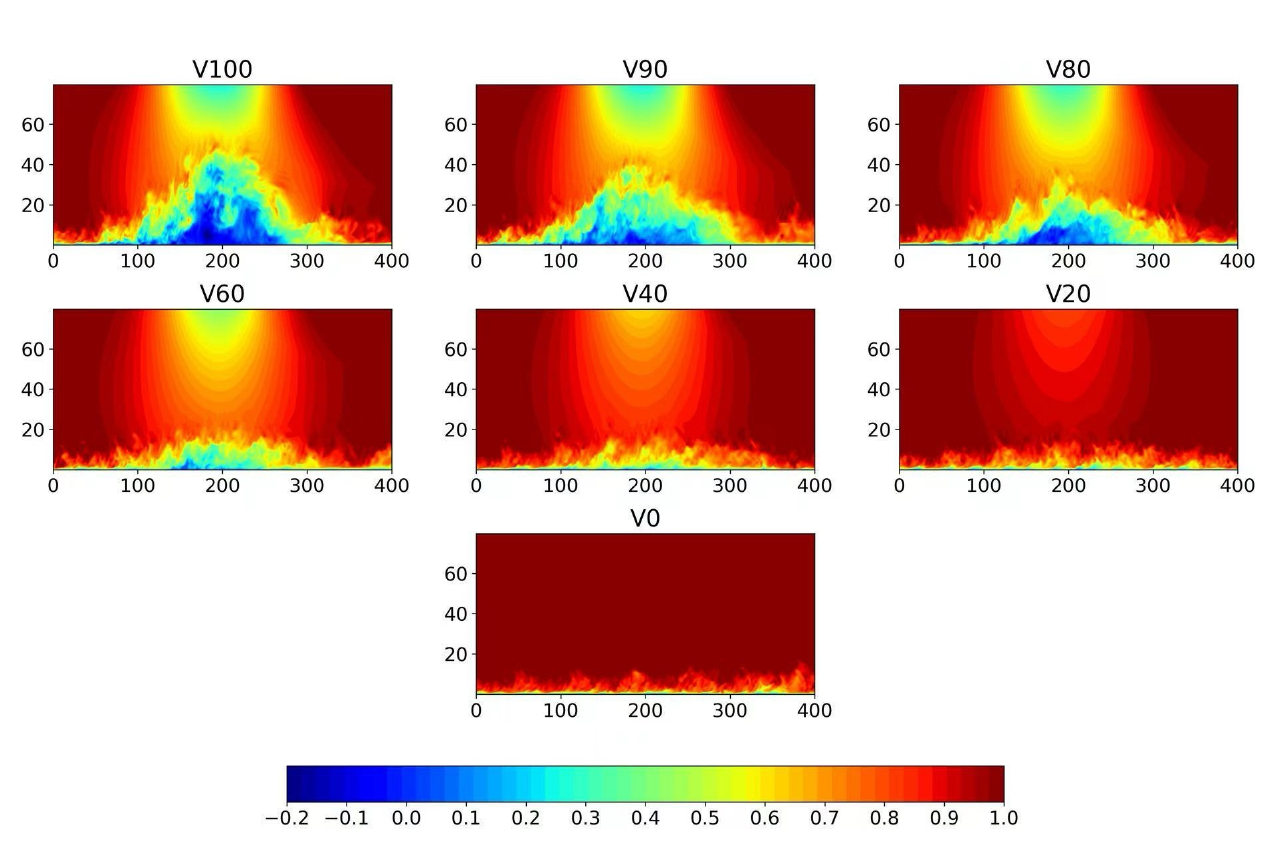}
    \caption{Instantaneous streamwise velocity contours for varying pressure gradients.}
    \label{fig:case_velocity}
\end{figure}

\begin{figure}[H]
    \centering
    \includegraphics[width=0.9\textwidth]{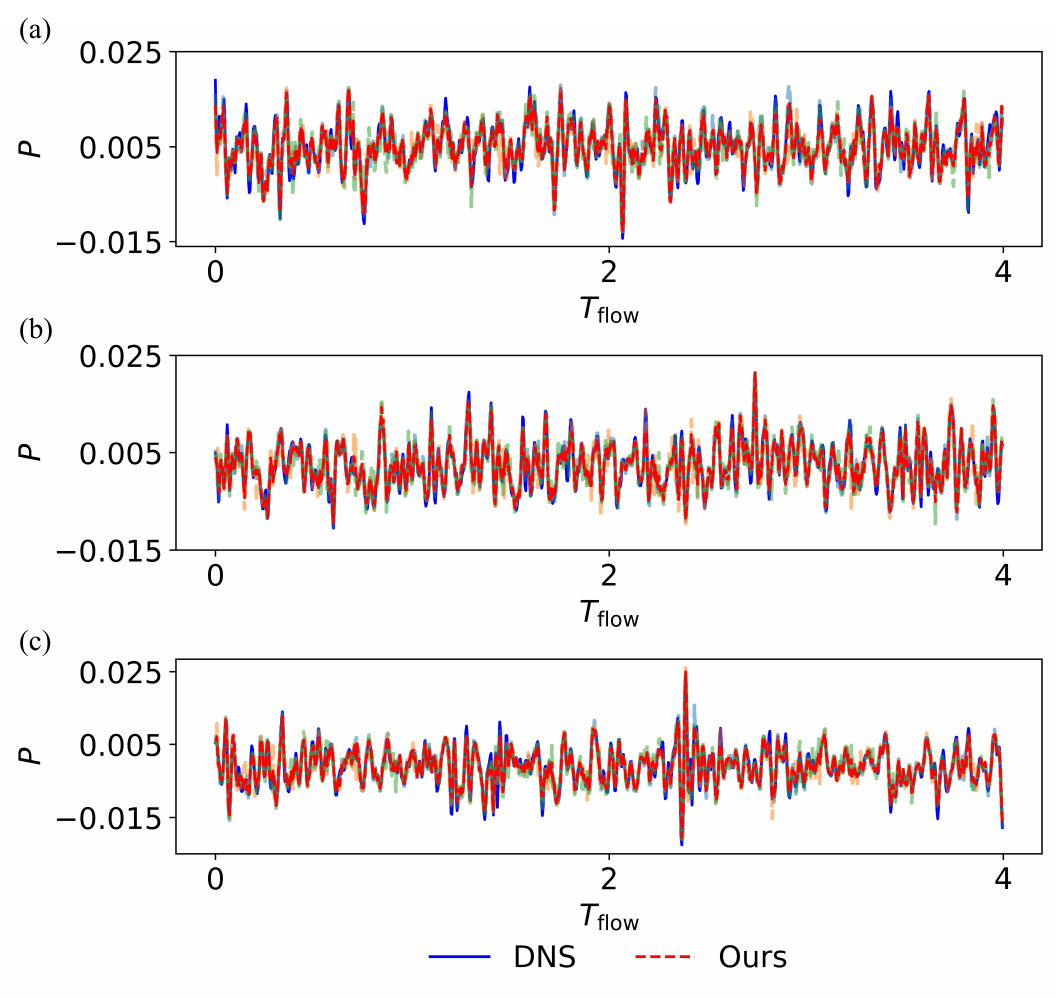}
    \caption{Time series of wall-pressure fluctuations at three unconditioned sensor locations for Case V0 (ZPG, $\beta = 0$), including three different sample realizations.}
    \label{fig:single_apg_v0_signal}
\end{figure}

\begin{figure}[H]
    \centering
    \includegraphics[width=0.9\textwidth]{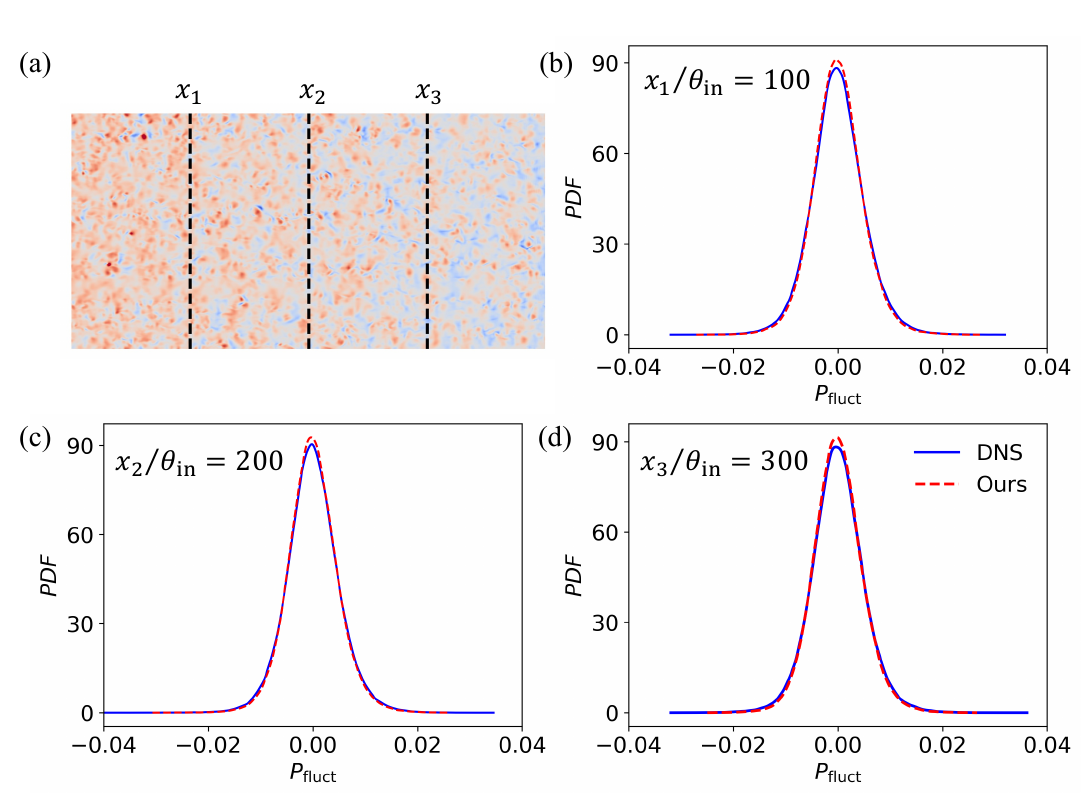}
    \caption{Probability density functions (PDFs) of wall-pressure fluctuations at three streamwise locations for Case V0 (ZPG, $\beta = 0$): (a) diagram showing the three locations; (b) PDF at $x_1/\theta_{\mathrm{in}} = 100$; (b) PDF at $x_2/\theta_{\mathrm{in}} = 200$; (c) PDF at $x_3/\theta_{\mathrm{in}} = 300$.}
    \label{fig:single_apg_v0_pdf}
\end{figure}

\begin{figure}[H]
    \centering
    \includegraphics[width=0.9\textwidth]{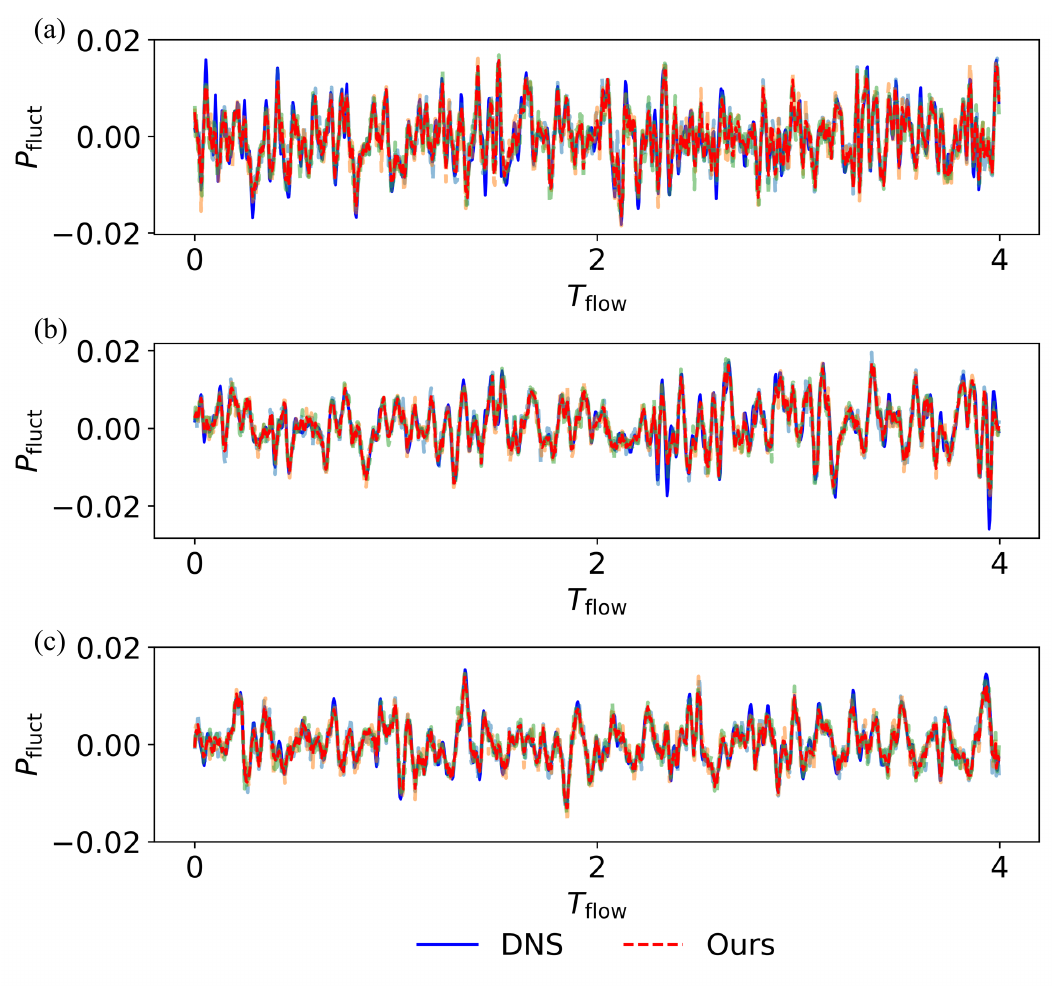}
    \caption{Time series of wall-pressure fluctuations at three unconditioned sensor locations for Case V40 (APG/FPG, $\beta = 6.92$), including three different sample realizations.}
    \label{fig:single_apg_v40_signal}
\end{figure}

\begin{figure}[H]
    \centering
    \includegraphics[width=0.9\textwidth]{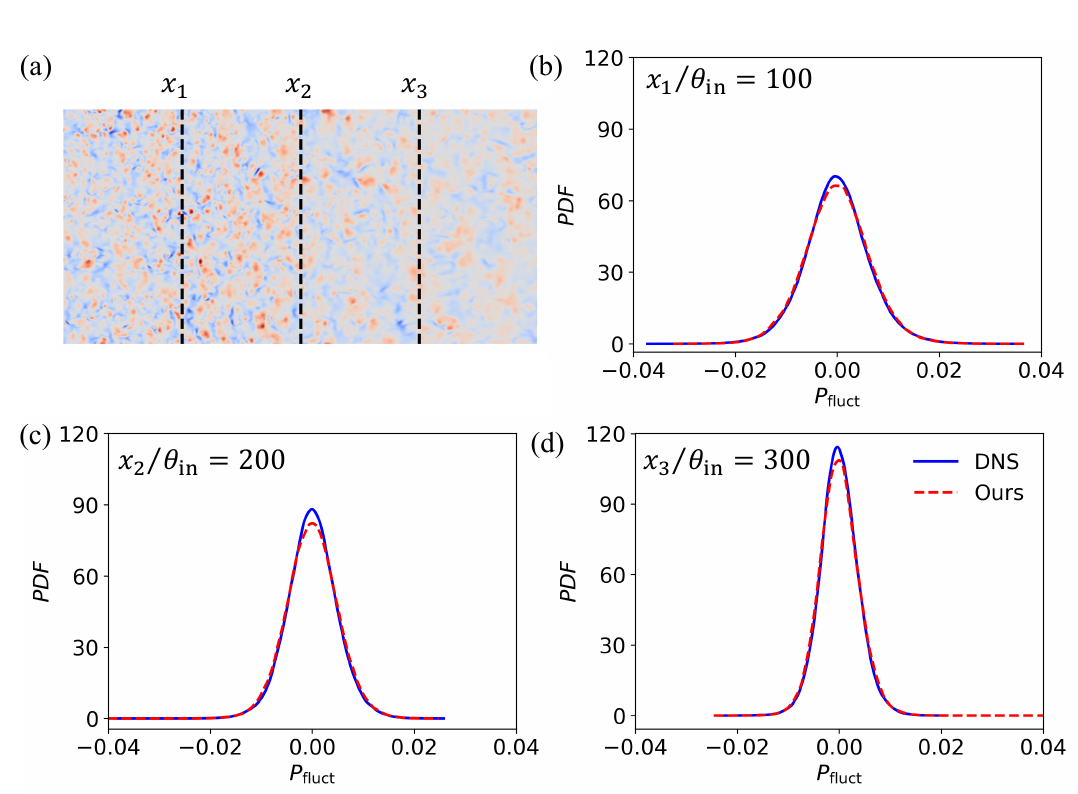}
    \caption{Probability density functions (PDFs) of wall-pressure fluctuations at three streamwise locations for Case V40 (APG/FPG, $\beta = 6.92$): (a) diagram showing the three locations; (b) PDF at $x_1/\theta_{\mathrm{in}} = 100$; (b) PDF at $x_2/\theta_{\mathrm{in}} = 200$; (c) PDF at $x_3/\theta_{\mathrm{in}} = 300$.}
    \label{fig:single_apg_v40_pdf}
\end{figure}

\begin{figure}[H]
    \centering
    \includegraphics[width=1.0\textwidth]{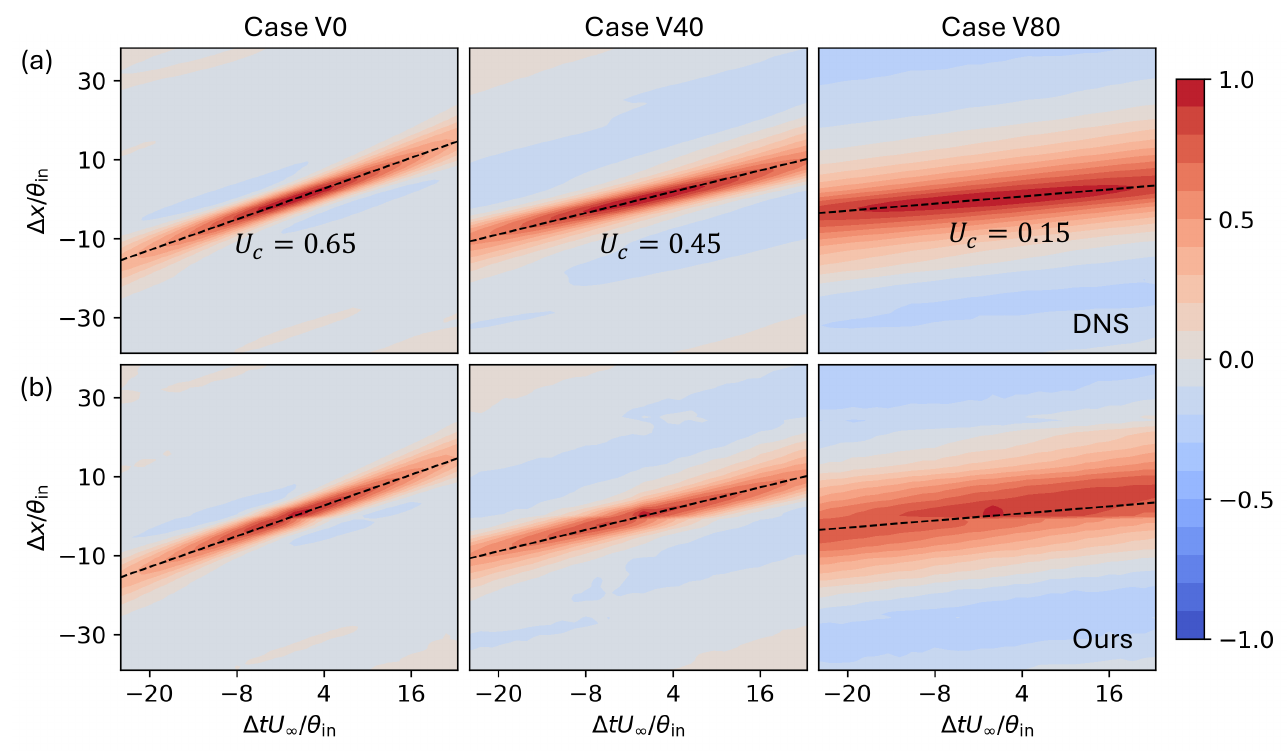}
    \caption{Contours of the streamwise space-time correlation coefficient at the streamwise location $x/ \theta_{\mathrm{in}} = 200$ for training ZPG and APG/FPG cases: (a) DNS results; (b) reconstructed field. Slopes of the dashed lines indicate the convection velocities $U_c$.}
    \label{fig:training_coe_x200}
\end{figure}

\begin{figure}[H]
    \centering
    \includegraphics[width=1.0\textwidth]{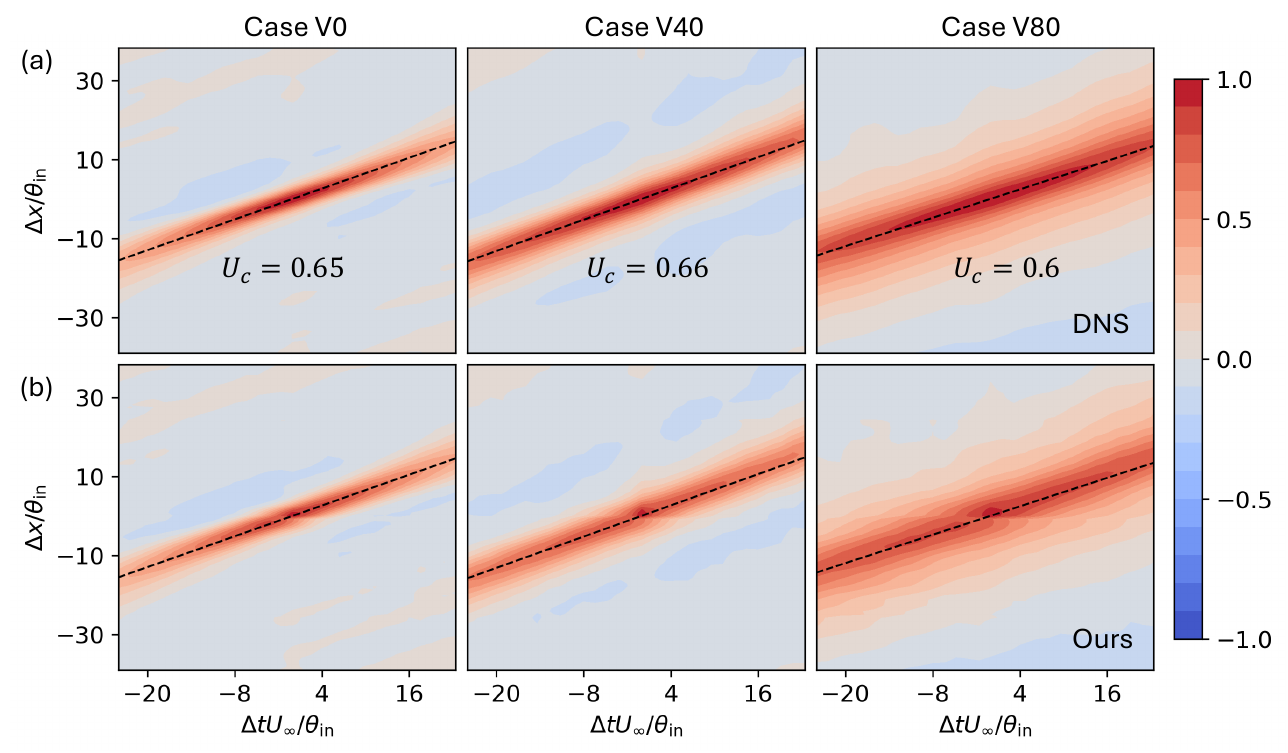}
    \caption{Contours of the streamwise space-time correlation coefficient at the streamwise location $x/ \theta_{\mathrm{in}} = 300$ for training ZPG and APG/FPG cases: (a) DNS results; (b) reconstructed field. Slopes of the dashed lines indicate the convection velocities $U_c$.}
    \label{fig:training_coe_x300}
\end{figure}

\begin{figure}[H]
    \centering
    \includegraphics[width=1.0\textwidth]{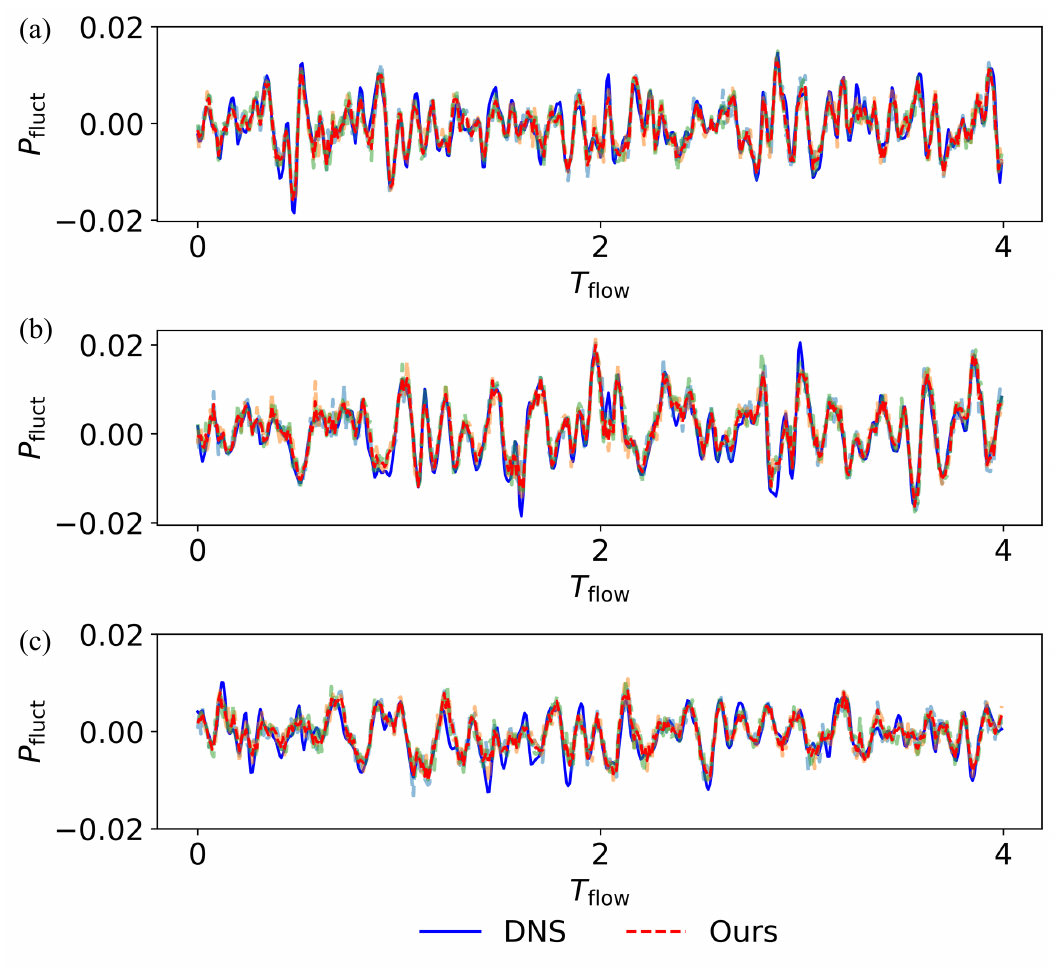}
    \caption{Time series of wall-pressure fluctuations at three unconditioned sensor locations for extrapolated Case V20 (APG/FPG, $\beta = 2.14$), including three different sample realizations.}
    \label{fig:single_apg_v20_signal}
\end{figure}

\begin{figure}[H]
    \centering
    \includegraphics[width=1.0\textwidth]{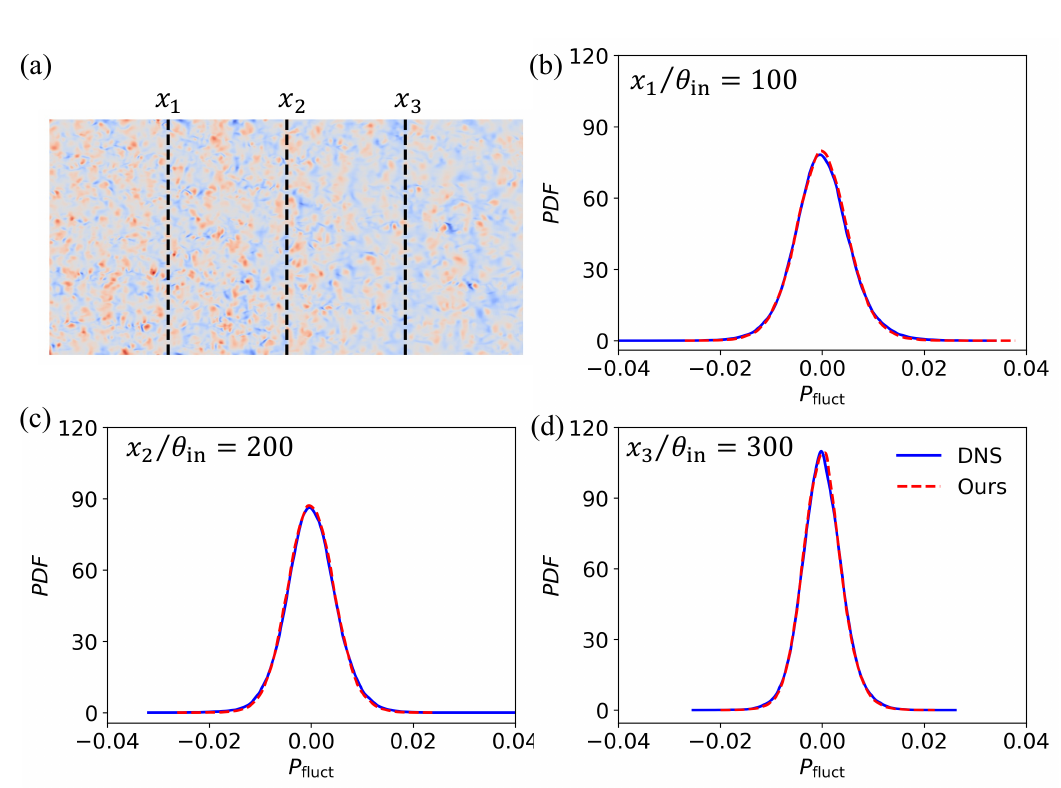}
    \caption{Probability density functions (PDFs) of wall-pressure fluctuations at three streamwise locations for extrapolated Case V20 (APG/FPG, $\beta = 2.14$): (a) diagram showing the three locations; (b) PDF at $x_1/\theta_{\mathrm{in}} = 100$; (b) PDF at $x_2/\theta_{\mathrm{in}} = 200$; (c) PDF at $x_3/\theta_{\mathrm{in}} = 300$.}
    \label{fig:single_apg_v20_pdf}
\end{figure}


\begin{figure}[H]
    \centering
    \includegraphics[width=1.0\textwidth]{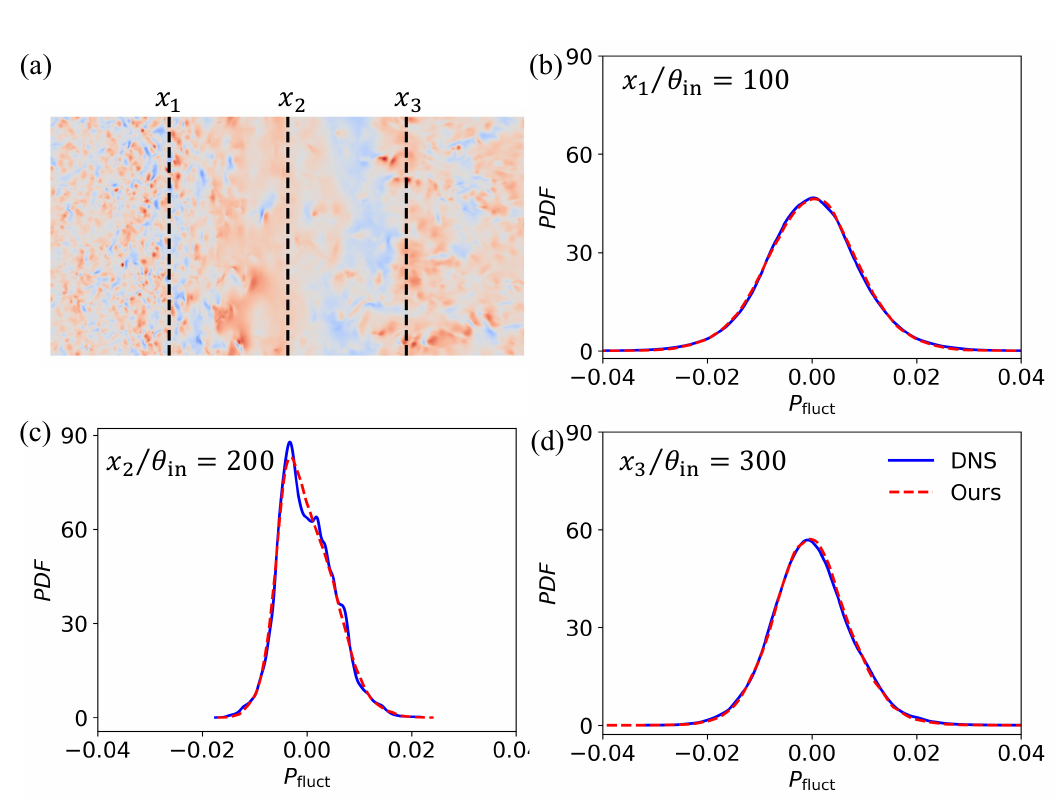}
    \caption{Probability density functions (PDFs) of wall-pressure fluctuations at three streamwise locations for extrapolated Case V90 (APG/FPG, $\beta = 90.19$): (a) diagram showing the three locations; (b) PDF near the separation point $x_1/\theta_{\mathrm{in}} = 100$; (b) PDF near the bubble center $x_2/\theta_{\mathrm{in}} = 200$; (c) PDF near the reattachment point $x_3/\theta_{\mathrm{in}} = 300$.}
    \label{fig:single_apg_v90_pdf}
\end{figure}

\begin{figure}[H]
    \centering
    \includegraphics[width=1.0\textwidth]{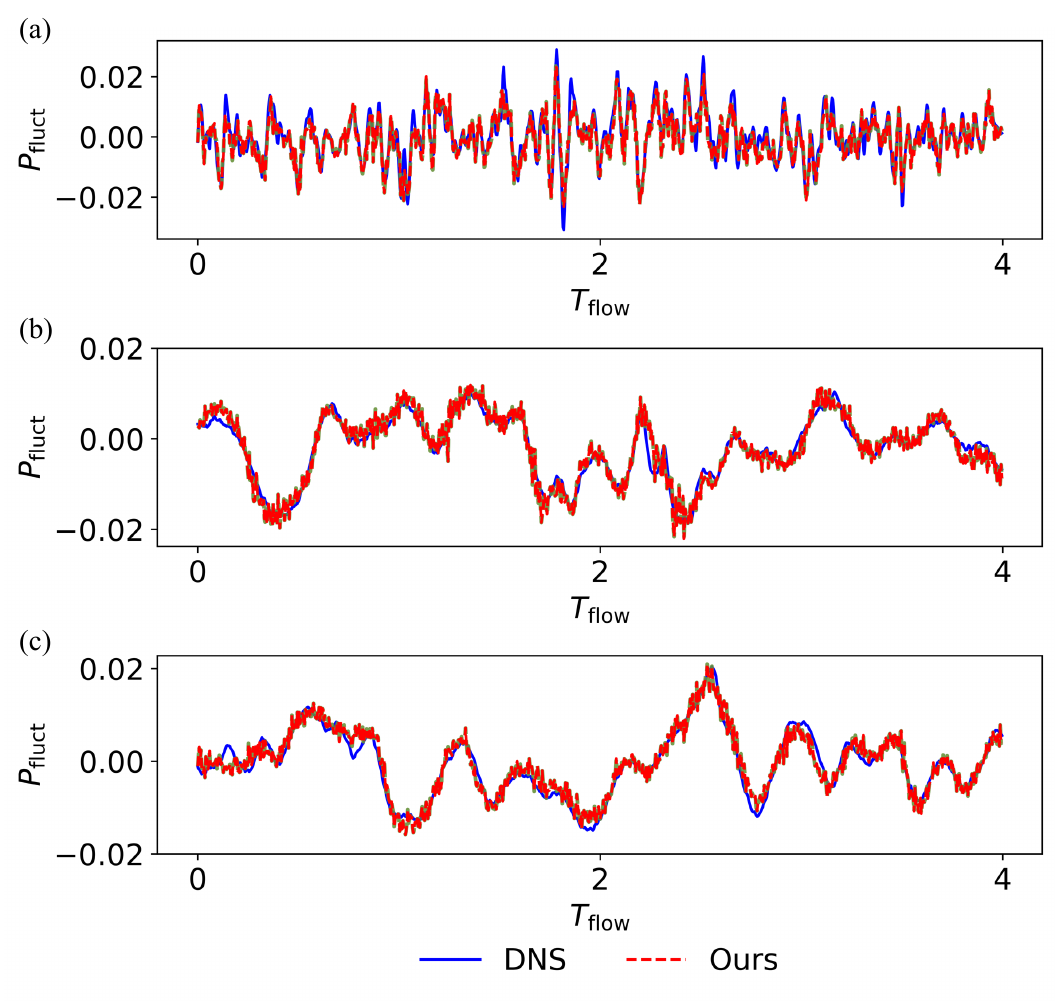}
    \caption{Time series of wall-pressure fluctuations at three unconditioned sensor locations for extrapolated Case V90 (APG/FPG, $\beta = 90.19$), including three different sample realizations.}
    \label{fig:single_apg_v90_signal}
\end{figure}

\begin{figure}[H]
    \centering
    \includegraphics[width=1.0\textwidth]{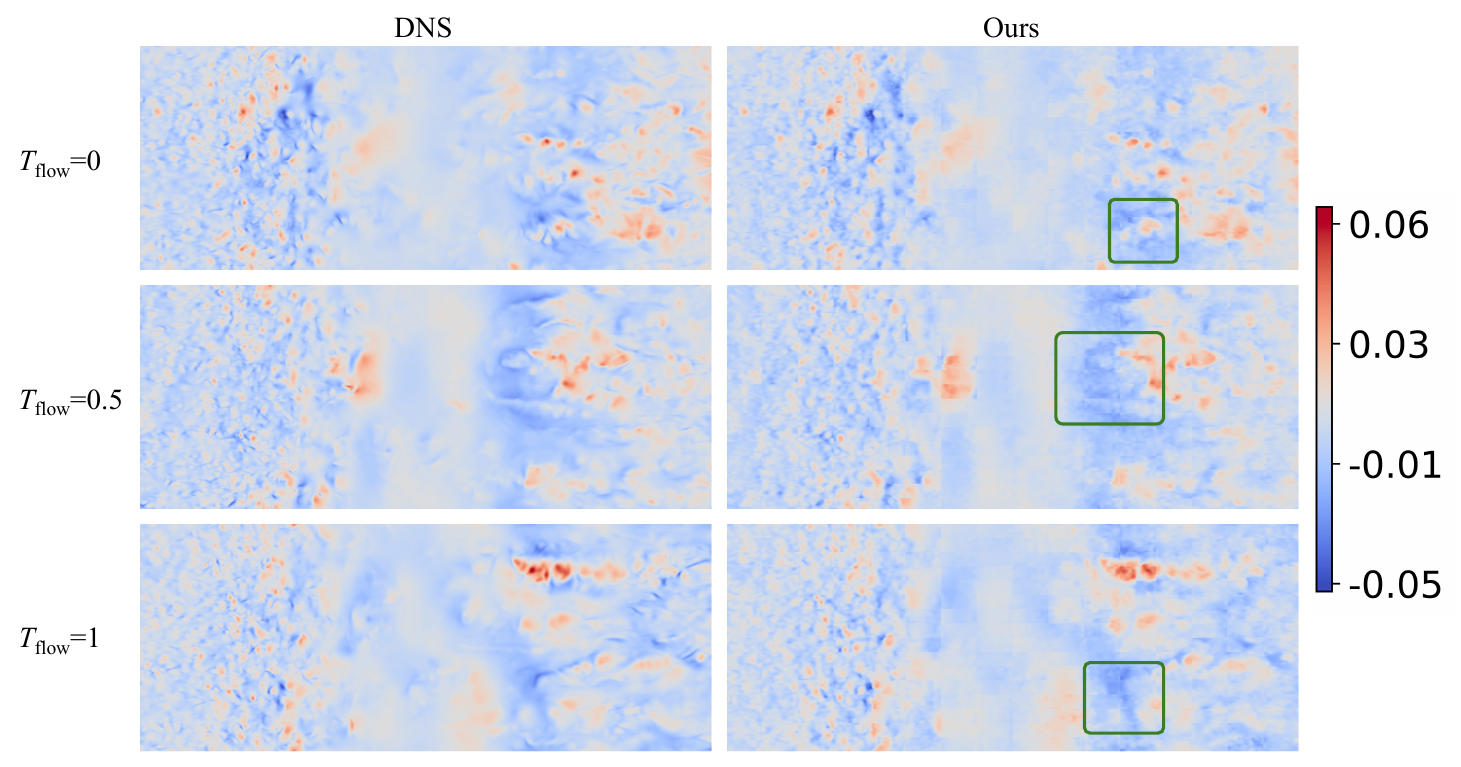}
    \caption{Reconstructed wall-pressure fluctuations for extrapolated Case V100 (APG/FPG, $\beta = 302.55$), conditioned on mean profiles and sensor arrays.}
    \label{fig:apg_v100_contour}
\end{figure}

\begin{figure}[H]
    \centering
    \includegraphics[width=1.0\textwidth]{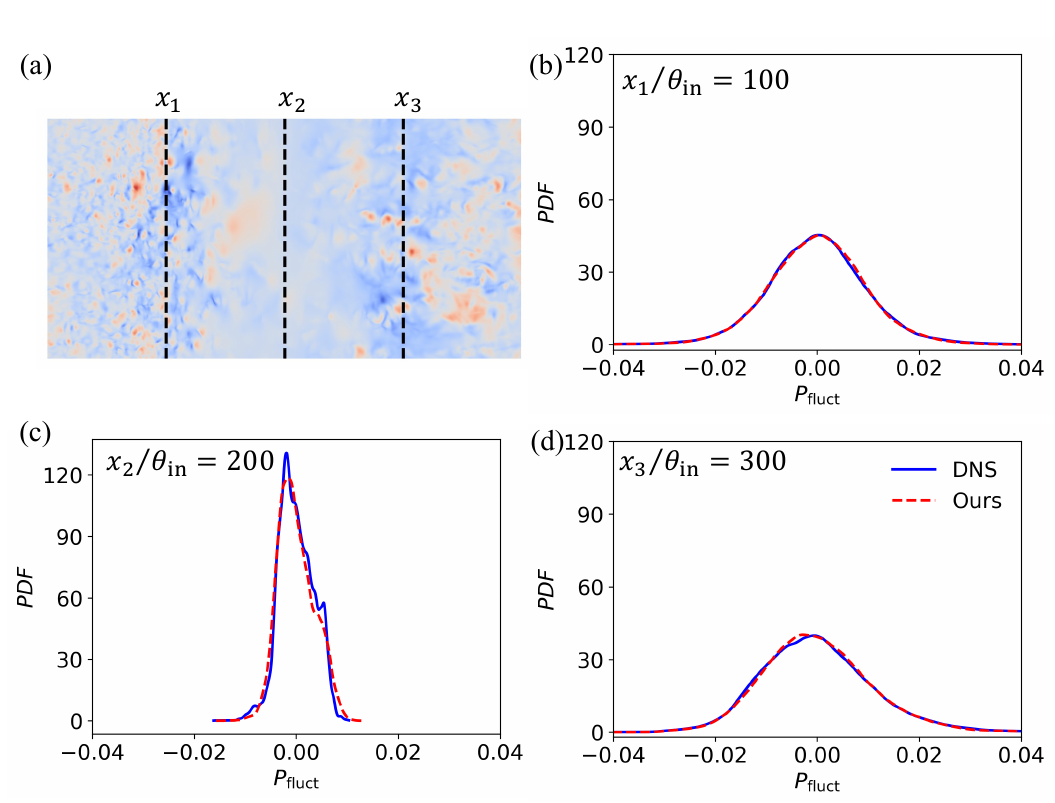}
    \caption{Probability density functions (PDFs) of wall-pressure fluctuations at three streamwise locations for extrapolated Case V100 (APG/FPG, $\beta = 302.55$): (a) diagram showing the three locations; (b) PDF near the separation point $x_1/\theta_{\mathrm{in}} = 100$; (b) PDF near the bubble center $x_2/\theta_{\mathrm{in}} = 200$; (c) PDF near the reattachment point $x_3/\theta_{\mathrm{in}} = 300$.}
    \label{fig:single_apg_v100_pdf}
\end{figure}

\begin{figure}[H]
    \centering
    \includegraphics[width=1.0\textwidth]{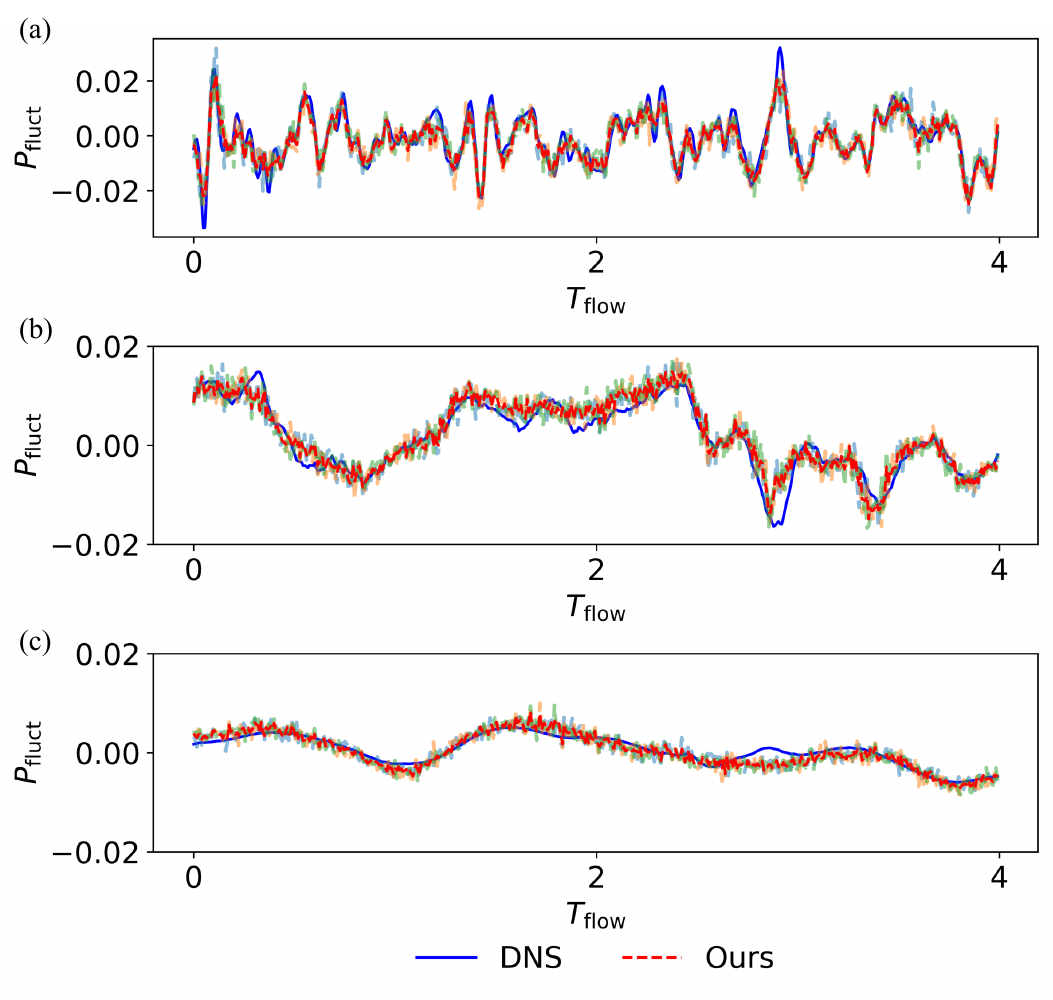}
    \caption{Time series of wall-pressure fluctuations at three unconditioned sensor locations for extrapolated Case V100 (APG/FPG, $\beta = 302.55$), including three different sample realizations.}
    \label{fig:single_apg_v100_signal}
\end{figure}

\begin{figure}[H]
    \centering
    \includegraphics[width=1.0\textwidth]{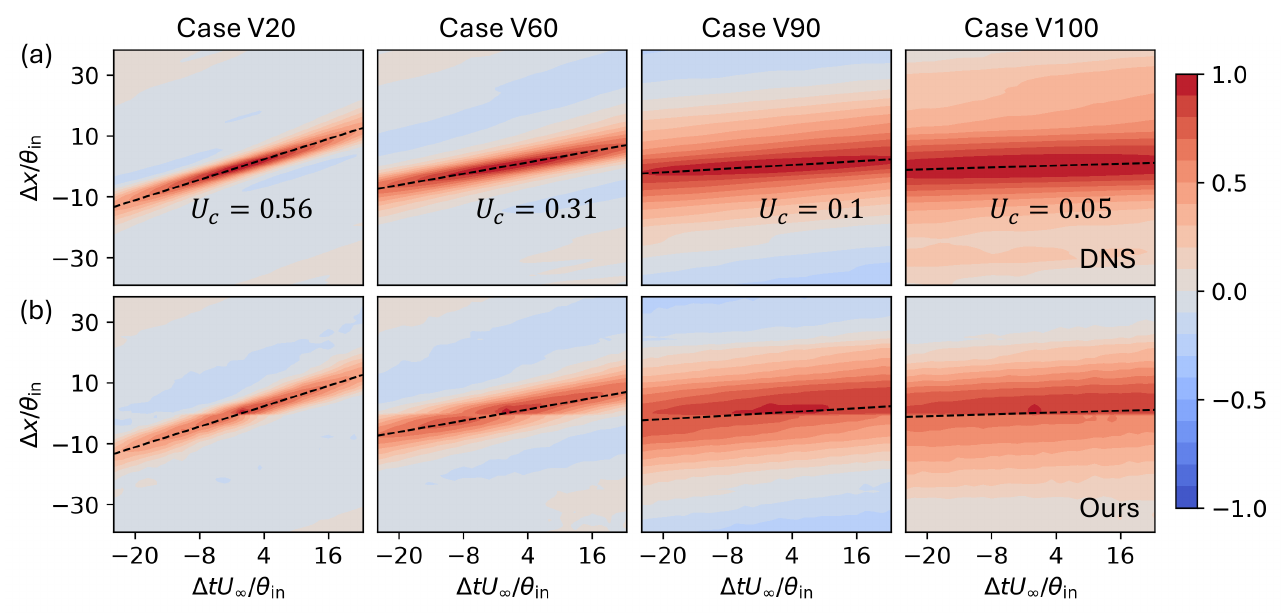}
    \caption{Contours of the streamwise space-time correlation coefficient at the streamwise location $x/ \theta_{\mathrm{in}} = 200$ for unseen APG/FPG cases: (a) DNS results; (b) reconstructed field. Slopes of the dashed lines indicate the convection velocities $U_c$.}
    \label{fig:testing_coe_x200}
\end{figure}

\begin{figure}[H]
    \centering
    \includegraphics[width=1.0\textwidth]{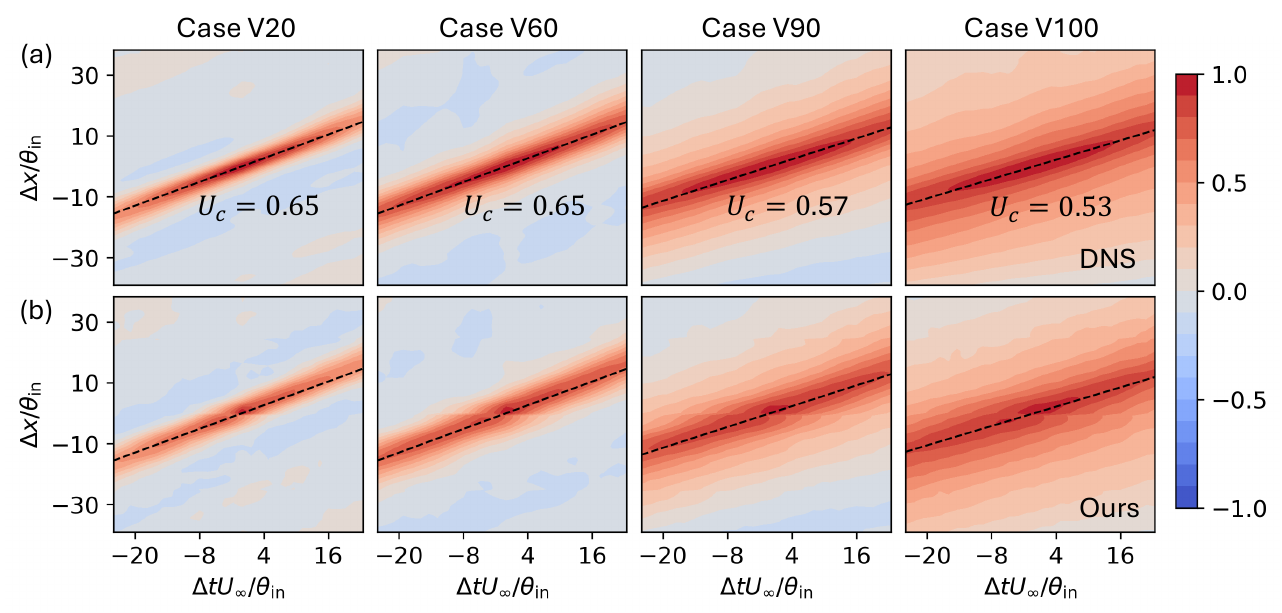}
    \caption{Contours of the streamwise space-time correlation coefficient at the streamwise location $x/ \theta_{\mathrm{in}} = 300$ for unseen APG/FPG cases: (a) DNS results; (b) reconstructed field. Slopes of the dashed lines indicate the convection velocities $U_c$.}
    \label{fig:testing_coe_x300}
\end{figure}

\begin{figure}[H]
    \centering
    \includegraphics[width=1.0\textwidth]{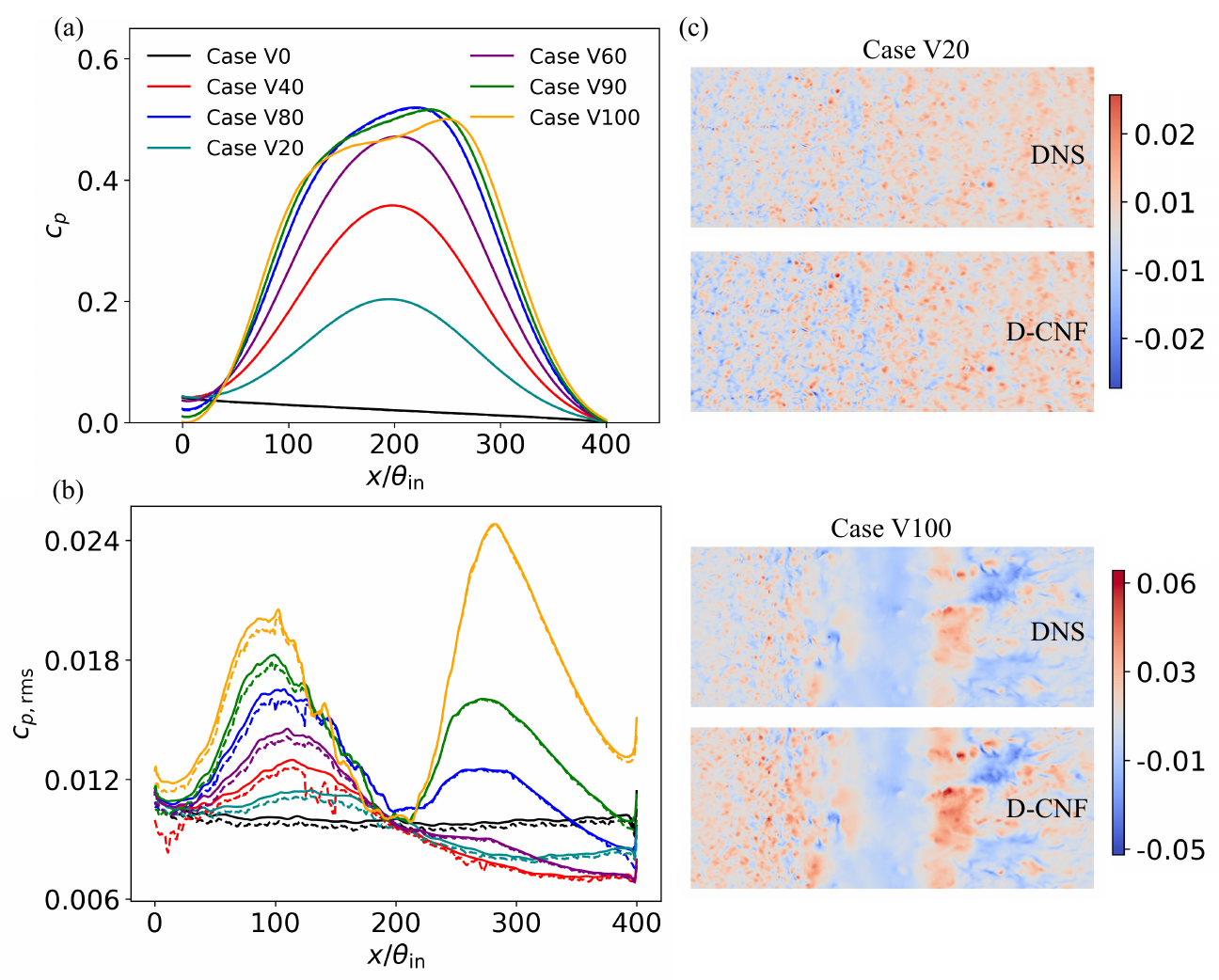}
    \caption{Results from the D-CNF applied to testing snapshots corresponding to $8-12$ flow-through times: (a) mean pressure coefficient $C_p$; (b) root-mean-square of pressure fluctuations $C_{p, \mathrm{rms}}$; and (c) selected instantaneous wall-pressure contours from DNS and decoded by the D-CNF. Solid lines represent DNS results, while dashed lines denote the results obtained from the D-CNF.}
    \label{fig:testing_cnf}
\end{figure}


\bibliographystyle{elsarticle-num}
\bibliography{ref,own-ref}

\end{document}